\documentclass[11pt, reqno]{article}
\usepackage{jheppub}
\usepackage{diagbox}
\usepackage{graphicx}
\usepackage{amsmath}
\usepackage{amssymb}
\usepackage{amsthm}
\usepackage{mathtools}
\usepackage{tikz}
\usepackage{tensor}
\usepackage{tikz-network}
\usepackage{mathrsfs}
\usepackage{epsfig}
\usepackage{dcolumn}
\usepackage{bm}
\usepackage{multirow}
\usepackage{natbib}
\usepackage{braket}
\usepackage{amsfonts}
\usepackage{dsfont}
\usepackage{hyperref}
\usepackage{xcolor}
\usepackage{subfigure}
\usepackage{paratensor}
\usepackage{paraKDHlattice}
\usepackage[thicklines]{cancel}
\usepackage{booktabs} % For professional-looking horizontal lines

\newcommand{\bea}{\begin{eqnarray}}
	\newcommand{\eea}{\end{eqnarray}}
\newcommand{\bean}{\begin{eqnarray*}}
	\newcommand{\eean}{\end{eqnarray*}}
\newcommand{\nn}{\nonumber \\}

\def\newline{{\hspace{15pt}}}

\def\abs#1{\left| #1\right|}
\def\braket#1{\left\langle #1 \right\rangle}
\def\bra#1{\left\langle #1\right|}
\def\ket#1{\left| #1\right\rangle}

\def\det{\mathop{\rm det}}

\def\eref#1{(\ref{#1})}

\def\co{\,,}
\def\ed{\,.}

\newcommand{\parall}[2]{{#1}\ /\kern -0.8em / \  {#2}}

\title{\boldmath Note on the $q=2$ $R$-para-fermionic SYK model  }

\author[a,b,c,d,e]{Tingfei Li}

\affiliation[a]{College of Physics Science and Technology, Hebei University, Baoding 071002, China}
\affiliation[b]{Hebei Key Laboratory of High-precision Computation and Application of Quantum Field Theory, Baoding, 071002, China}
\affiliation[c]{Hebei Research Center of the Basic Discipline for Computational Physics, Baoding, 071002, China}
\affiliation[d]{Zhejiang Institute of Modern Physics, Zhejiang University, Hangzhou, 310027, P. R. China }
\affiliation[e]{Kavli Institute for Theoretical Sciences (KITS), University of Chinese Academy of Sciences, Beijing 100190, China}

\emailAdd{tfli@zju.edu.cn}
\abstract{We investigate the $q=2$ SYK model with $R$-para-particles ($R$-PSYK$_2$), analyzing its thermodynamics and spectral form factor (SFF) using random matrix theory. The Hamiltonian is quadratic, with coupling coefficients randomly drawn from the Gaussian Unitary Ensemble (GUE). The model displays self-averaging behavior and exhibits an exponential ramp in its SFF dynamics: $\mathcal{K}(t) \sim e^{C_0 t}$. The growth rate $C_0$ tends toward either a constant or infinity in the $N\to \infty$ limit, depending on  specific statistics of the model. These results provide novel perspectives on quantum systems with unconventional statistics.  }

\begin{document}
	\maketitle
	\flushbottom

	\section{Introduction}
	\paragraph{Motivation} It is well known that fundamental particles are classified into two types: bosons and fermions, with their exchange statistics governed by commutation and anti-commutation relations. When we extend beyond fundamental particles in three-dimensional space, additional statistical possibilities emerge beyond these two cases. A prominent example are anyons in two-dimensional space~\cite{1977NCimB_anyon,PhysRevLett.49.957,Wilczek1990,Nayak_2008,2008AnPhy}, which play crucial roles in topological quantum computation and quantum phases of matter. 
	
	Beyond anyons, paraparticles \cite{PhysRev.90.270,Araki1961,Greenberg1965,Landshoff1967,Druehl1970,Robert1970,Baker2015} constitute another important generalization of ordinary exchange statistics that can be consistently defined in any spatial dimension. First studied by Green in 1953 \cite{PhysRev.90.270}, paraparticles represent a broader class of statistics that includes both bosons and fermions as special cases. Here we mainly consider \textit{$R$-para-particles} \footnote{Here, since parafermion has already been used extensively in the literature, we use a different name to distinguish the paraparticles with other definitions.} which are defined by an $R$-matrix in its commutation relations \cite{Wang_2025,wang2025parastatisticssecretcommunicationchallenge}. As demonstrated in \cite{Wang_2025,wang2025parastatisticssecretcommunicationchallenge}, $R$-para-particles can emerge as quasi-particles in condensed matter systems, suggesting their potential fundamental nature. However, recent arguments \cite{mekonnen2025invariancequantumpermutationsrules} have challenged this view, proposing that only bosons and fermions can exist as fundamental particles in nature.
	
	Nevertheless, $R$-para-particles can be engineered in condensed matter systems or quantum computing platforms, making their study both meaningful and practical. Following conventional notation, we represent the creation and annihilation operators for $R$-para-particles as
	\begin{align}
		\hat{\psi}_{i,a}^{+},\hat{\psi}_{i,a}^{-}
	\end{align}
	where $i = 1,2,\ldots,N$ denotes the site index and $a = 1,2,\ldots,m$ represents the flavor index. These operators obey generalized exchange statistics described by~\footnote{From now on, we write $\psi$ without the hat, since no confusion should arise.}
	\begin{equation}
		\begin{aligned}
			{\psi}_{l,a}^-{\psi}_{j,b}^+ &= \sum_{cd} R^{ac}_{bd} {\psi}_{j,c}^+ {\psi}_{l,d}^- + \delta_{ab} \delta_{lj}, \\
			{\psi}_{l,a}^+ {\psi}_{j,b}^+ &= \sum_{cd} R^{cd}_{ab} {\psi}_{j,c}^+ {\psi}_{l,d}^+, \\
			{\psi}_{l,a}^- {\psi}_{j,b}^- &= \sum_{cd} R^{ba}_{dc} {\psi}_{j,c}^- {\psi}_{l,d}^-\ed
		\end{aligned}
	\end{equation}
	In this formulation, each type of parastatistics is labeled by a four-index tensor $R^{ab}_{cd}$ satisfying  the constant
	Yang-Baxter equation (YBE) \cite{Turaev1988TheYE,10.1063/1.529485,etingof1998settheoreticalsolutionsquantumyangbaxter} %
	\begin{equation}\label{eq:YBE}
		\begin{tikzpicture}[baseline={([yshift=-.8ex]current bounding box.center)}, scale=0.5]
			\Rmatrix{0}{\AL}{R}
			\Rmatrix{0}{-\AL}{R}
			\node  at (-\AL,2.5*\AL) {\footnotesize $a$};
			\node  at (\AL,2.5*\AL) {\footnotesize $b$};
			\node  at (-\AL,-2.5*\AL) {\footnotesize $c$};
			\node  at (\AL,-2.5*\AL) {\footnotesize $d$};
		\end{tikzpicture}=
		\begin{tikzpicture}[baseline={([yshift=-.8ex]current bounding box.center)}, scale=0.5]
			\draw[thick] (-\AL,-2*\AL) -- (-\AL,2*\AL);
			\draw[thick] (\AL,-2*\AL) -- (\AL,2*\AL);
			\node  at (-\AL,2.5*\AL) {\footnotesize $a$};
			\node  at (\AL,2.5*\AL) {\footnotesize $b$};
			\node  at (-\AL,-2.5*\AL) {\footnotesize $c$};
			\node  at (\AL,-2.5*\AL) {\footnotesize $d$};
			\node  at (-1.5*\AL,0*\AL) {\footnotesize $\delta$};
			\node  at (1.5*\AL,0*\AL) {\footnotesize $\delta$};
		\end{tikzpicture}~,~~~
		\begin{tikzpicture}[baseline={([yshift=-.8ex]current bounding box.center)}, scale=0.5]
			\Rmatrix{-\AL}{2*\AL}{R}
			\Rmatrix{\AL}{0}{R}
			\Rmatrix{-\AL}{-2*\AL}{R}
			\draw[thick] (-2*\AL,-\AL) -- (-2*\AL,\AL);
			\draw[thick] (2*\AL,\AL) -- (2*\AL,3*\AL);
			\draw[thick] (2*\AL,-\AL) -- (2*\AL,-3*\AL);
			\node  at (-2*\AL,3.5*\AL) {\footnotesize $a$};
			\node  at (0*\AL,3.5*\AL) {\footnotesize $b$};
			\node  at (2*\AL,3.5*\AL) {\footnotesize $c$};
			\node  at (-2*\AL,-3.7*\AL) {\footnotesize $d$};
			\node  at (0*\AL,-3.7*\AL) {\footnotesize $e$};
			\node  at (2*\AL,-3.7*\AL) {\footnotesize $f$};
		\end{tikzpicture}
		=\begin{tikzpicture}[baseline={([yshift=-.8ex]current bounding box.center)}, scale=0.5]
			\Rmatrix{\AL}{2*\AL}{R}
			\Rmatrix{-\AL}{0}{R}
			\Rmatrix{\AL}{-2*\AL}{R}
			\draw[thick] (2*\AL,-\AL) -- (2*\AL,\AL);
			\draw[thick] (-2*\AL,\AL) -- (-2*\AL,3*\AL);
			\draw[thick] (-2*\AL,-\AL) -- (-2*\AL,-3*\AL);
			\node  at (-2*\AL,3.5*\AL) {\footnotesize $a$};
			\node  at (0*\AL,3.5*\AL) {\footnotesize $b$};
			\node  at (2*\AL,3.5*\AL) {\footnotesize $c$};
			\node  at (-2*\AL,-3.7*\AL) {\footnotesize $d$};
			\node  at (0*\AL,-3.7*\AL) {\footnotesize $e$};
			\node  at (2*\AL,-3.7*\AL) {\footnotesize $f$};
		\end{tikzpicture},
	\end{equation}
	where $R^{ab}_{cd}=\!\!\begin{tikzpicture}[baseline={([yshift=-.6ex]current bounding box.center)}, scale=0.45]
		\Rmatrix{0}{0}{R}
		\node  at (-1.5*\AL,\AL) {\footnotesize $a$};
		\node  at (1.5*\AL,\AL) {\footnotesize $b$};
		\node  at (-1.5*\AL,-\AL) {\footnotesize $c$};
		\node  at (1.5*\AL,-\AL) {\footnotesize $d$};
	\end{tikzpicture}$,
	and a line segment represents a Kronecker $\delta$ function. Note that the $R$-tensor is site-independent, which represents a non-trivial constraint.  For generic $R$, this makes the operators $\psi_{i,a}^{\pm}$ inherently non-local: a fundamental distinction between $R$-para-particles and conventional fermions/bosons. To maintain the physical interpretation of $\psi_{i,a}^{\pm}$ as creation and annihilation operators, we impose an additional condition
	\begin{align}\label{eq:unitary-condition}
		\sum_{a,b}R_{cd}^{ab}\left(R_{ef}^{ab}\right)^{*}=\delta_{ce}\delta_{df}\co	
	\end{align}
	so that $\psi_{i,a}^+ = (\psi_{i,a}^-)^\dagger$. We emphasize that the special cases $R^{ab}_{cd} = \pm \delta_{ad}\delta_{bc}$ recover standard bosonic ($+$) and fermionic ($-$) statistics, demonstrating that $R$-para-particles constitute a natural generalization of these fundamental particle types.

	Recently, a model called SYK (Sachdev-Ye-Kitaev) \cite{Kitaev-talks,Kitaev2015,Polchinski_2016,Maldacena_2016,Jevicki:2016bwu,jevicki2016bilocalholographysykmodel,Kitaev:2017awl} and its generalizations \cite{Witten:2016iux,Fu:2016vas,Klebanov:2016xxf,Gross:2016kjj,Berkooz:2016cvq,Murugan:2017eto,Peng:2017kro,Peng:2016mxj,Gu:2019jub,Berkooz:2018jqr} have been the focus of extensive research. This model consists of $q$-body interactions among fermions with Gaussian random couplings. Its many remarkable properties make it a useful model for studying quantum chaos \cite{Maldacena_2016chaos,Parker:2018yvk,Gu:2016oyy,Roberts:2018mnp}, black holes \cite{Maldacena:2018lmt,Maldacena:2017axo,Nayak:2018qej}, holography \cite{Saad:2018bqo,Sarosi:2017ykf,Jevicki:2016bwu,Gross:2017hcz}, and quantum matter \cite{Davison:2016ngz,Hartnoll:2016apf,Song:2017pfw,Chowdhury:2021qpy}. Here, we briefly review its basic properties.
	
	The SYK model can be solved in the IR limit, the large $q$ limit, and the double-scaled limit \cite{Berkooz:2018jqr} for large $N$. In the IR limit, it acquires conformal symmetry, and the effective action can be approximated by the Schwarzian one, indicating its duality to 2D dilaton gravity. Its double-scaled limit has connections to dS holography \cite{Lin:2022rbf,Narovlansky:2023lfz,Susskind:2022bia}. Moreover, it exhibits exponentially decaying out-of-time-order correlators (OTOCs) with the maximal exponent $\frac{2\pi}{\beta}$ \cite{Maldacena_2016,Maldacena_2016chaos} and a linear ramp in its SFF, which serve as signatures of quantum chaos.
	
	We are interested in the properties of the SYK model constructed with $R$-para-particles, which we refer to as $R$-PSYK for simplicity.~\footnote{There are other ways to introduce paraparticle properties in the SYK model. For example, as shown in \cite{Peng:2018zap}, one can vary the parameter $\mu$ such that the solution to the Schwinger-Dyson equation interpolates between fermionic and bosonic behaviors, with the intermediate region corresponding to a para-SYK description.}  Given that the path integral method plays a crucial role in evaluating the SYK model, we now explore whether it still applies to $R$-PSYK. However, there is a key difference between ordinary particles and $R$-para-particles.~\footnote{Here we do not consider the trivial case where statistics returns back to fermions or bosons.} For nontrivial $R$, the ``creation'' and ``annihilation'' operators of $R$-para-particles are global, even if they can be written as matrix product operators (MPOs) of local spin operators by a significant generalization of the Jordan-Wigner transformation (JWT) \cite{Wang_2025}. The complicated exchange statistics make it difficult to derive a path integral formulation.~\footnote{A simpler alternative is to consider commuting SYK models, as explored in \cite{Gao:2023gta,Gao:2024lem}.} Although any local operator in a finite-dimensional Hilbert space can be expressed in terms of fermionic creation and annihilation operators—enabling a path integral formulation in principle, or alternatively, the use of coherent state path integrals—the Hamiltonian's nonlocality may make the resulting expression too cumbersome for practical applications.
	
	Beyond the path integral method, the statistics of random coupling coefficients may simplify the problem in certain limits. For example, in the double-scaled limit \cite{Berkooz:2018jqr}, ensemble-averaged moments $\mathbb{E}(\text{tr}H^k)$ can be calculated analytically, as exchanging two Hamiltonians in the trace only introduces a factor, simplifying the construction of the partition function and correlation functions. However, for $R$-PSYK, exchanging two operators leads to complex results, so we do not expect similar simplifications in the double-scaled limit.
	
	As a first step, it is natural to consider the simplest case: the non-interacting para-SYK model ($q=2$). While the SYK model is typically studied for $q>2$ (since a simple transformation can reduce the $q=2$ case to a free model that might appear trivial), the disorder in the couplings can still produce interesting physics. Although the $q=2$ model loses some key features of the standard SYK, the SYK$_2$ model retains several nontrivial characteristics. Most notably, it displays an exponential ramp in its spectral form factor (SFF), a feature numerically verified in \cite{Lau_2019} that signals chaotic behavior.
	Beyond the exact solvability of its Green's function, the simplicity of SYK$_2$ enables a precise analysis of its level statistics. Its SFF can be understood using random matrix theory techniques \cite{Liao_2020} as well as the $G\Sigma$ path integral formalism \cite{Winer_2020,Chris_Lau_2021,legramandi2024manybodyspectraltransitionslens}. Moreover, the tractability of SYK$_2$ allows for exact verification of eigenstate thermalization (ETH) \cite{syk2eth2021} and detailed studies of eigenstate entanglement \cite{Liu_2018,syk2entangle2021}. Further analysis of the SFF for SYK$_2$ can be found in \cite{legramandi2024momentsspectralformfactor}. For a comprehensive review of the $q=2$ SYK model we refer to \cite{pethybridge2024notescomplexq2syk}.

	\begin{table}[ht]\label{table:six-cases}
		\centering
		\vspace{0.5cm}
		\begin{tabular}{l|ll} % Vertical line added after the first column
			\toprule
			& \textbf{$R$-para-fermions} & \textbf{Dual $R$-para-bosons} \\
			\midrule
			Ordinary & $(1+x)^m$ & $(1-x)^{-m}$ \\
			Example A & $1+mx$ & $(1-mx)^{-1}$ \\
			Example B & $1+mx+x^2$ & $(1-mx+x^2)^{-1}$ \\
			\bottomrule
		\end{tabular}
		\caption{The single-mode partition functions of the three types of $R$-para-fermions and their dual $R$-para-bosons are studied in this paper. The function $z_R(x) = 1 + m x$ is obtained when $R^{ab}_{cd} = -\delta_{ac} \delta_{bd}$, whereas $z_R(x) = 1 + m x + x^2$ corresponds to $R^{ab}_{cd} = \lambda_{ab} \xi_{cd} - \delta_{ac} \delta_{bd}$. Here, the matrices $\lambda$ and $\xi$ satisfy the conditions $\lambda \xi \lambda^T \xi^T = \mathbf{1}$ and $\operatorname{Tr}(\lambda \xi^T) = 2$. }
	\end{table}
	
	\paragraph{The model} In this paper, we mainly study the $q=2$ SYK model constructed by $R$-para-particles, which we refer to as $R$-PSYK$_2$, with the Hamiltonian
	\begin{align}
		H = \sum_{a=1}^{m}\sum_{1\le i,j\le N}(h_{ij}-\mu \delta_{ij})\psi_{i,a}^{+}\psi_{j,a}^{-}
	\end{align}
	where $\mu$ is the chemical potential and $h_{ij}$ is a random matrix drawn from the Gaussian unitary ensemble (GUE).  The matrix elements are distributed as follows: diagonal elements $h_{ii} \sim \mathcal{N}(0,1/N)$, while off-diagonal elements have $\mathrm{Re}(h_{ij}) \sim \mathcal{N}(0,1/(2N))$ and $\mathrm{Im}(h_{ij}) \sim \mathcal{N}(0,1/(2N))$, where $\mathcal{N}(\mu,\sigma^2)$ denotes the normal distribution with mean $\mu$ and variance $\sigma^2$. The indices $i,j = 1,2,\ldots,N$ label the lattice sites, while $a = 1,2,\ldots,m$ represents the flavor index. Throughout this work, we maintain the unitary condition specified in Eq.~\eqref{eq:unitary-condition}, which ensures the relation $\psi_{i,a}^+ = (\psi_{i,a}^-)^\dagger$ holds consistently. 
	
	The model can be transformed into a free model via the substitution $\psi_{i,a}^- \to U_{ij}\widetilde{\psi}_{j,a}^-$, where $U_{ij}$ diagonalizes $h_{ij}$. The new operators satisfy the same commutation relations, so we will not distinguish them. As in ordinary fermionic/bosonic systems, we interpret $\psi_{i,a}^{+}$ and $\psi_{i,a}^{-}$ as the ``creation'' and ``annihilation'' operators of $R$-para-particles. We define the vacuum state by $\psi_{i,a}^-\ket{0} = 0$ for all $i,a$, and construct $n$-particle states as $\psi_{i,a_1}^+\psi_{i,a_2}^+\ldots \psi_{i,a_n}^+\ket{0}$ for a given site. 
	
	We find that the thermodynamics and SFF of the model \textbf{only}  depend on the dimension of the $n$-particle Hilbert space (or the degeneracy of the $n$-th level), denoted by $d_n$ for $n = 0,1,2,\ldots$. The value of $d_n$ is determined by the choice of the $R$-matrix. This information can be encoded in the single-mode partition function
	\begin{align}
		z_R(x)\equiv \sum_{n=0}^\infty d_n x^n\ed
	\end{align}
	For ordinary fermions and bosons with $m$ flavors, the single-mode partition functions are given by  
	\begin{align}  
		z_{\text{Fermion}}(x) = (1 + x)^{m}, \quad  
		z_{\text{Boson}}(x) = (1 - x)^{-m}.  
	\end{align}  
	In this paper, we adopt a simplified terminology: \textit{$R$-para-fermions} refer to $R$-para-particles with a finite polynomial partition function $z_R(x) = \sum_{n=1}^L d_n x^n$ (where $L$ is finite), while \textit{$R$-para-bosons} describe $R$-para-particles with a fractional $z_R(x)$. 
	Since $R$ satisfies the Yang-Baxter equation \eqref{eq:YBE}, so does $-R$, leading to a duality between $R$-para-bosons and $R$-para-fermions:  
	\begin{align}  
		z_{R}(x)z_{-R}(-x) = 1\ed 
	\end{align}  
	When $R^{ab}_{cd} \neq \pm \delta_{ad}\delta_{bc}$ but $z_R(x) = (1 + x)^{m}$ or $(1 - x)^{-m}$, we say the $R$-PSYK$_2$ is \textit{trivial}, as it shares the same grand canonical partition function and SFF with the fermionic/bosonic $\mathrm{SYK}_2$.~\footnote{While these cases are not entirely trivial—they may yield distinct correlation functions compared to conventional $\mathrm{SYK}_2$—here we focus primarily on their thermodynamics and SFF.}
	In this work, we study $R$-PSYK$_2$ with three distinct forms of $z_R(x)$ and their duals, as summarized in Table~\ref{table:six-cases}. Thus, Example~A ($m=1$) and Example~B ($m=2$) correspond to trivial $R$-PSYK$_2$. Notice that the unitary condition in Eq.~\eref{eq:unitary-condition} cannot be satisfied for Example B with \( m \ge 3 \) for any \(\lambda, \xi\). As a result, we never have \(\psi_{i,a}^+ = (\psi_{i,a}^-)^\dagger\) in this case. However, the calculation of the partition function and SFF depends only on \(d_n\), meaning the physical results in the paper remain valid. Nevertheless, care must be taken when computing correlation functions of \(\psi\).\footnote{We thank Zhiyuan Wang, an author of \cite{Wang_2025}, for pointing out this.}
	\paragraph{Main results}
	As discussed earlier, studying the $R$-PSYK model using the path integral method is not convenient. In this paper, we primarily analyze the model in the large $N$ limit using random matrix theory. Our focus is on the thermodynamics and the SFF of the model. Additionally, we examine the time evolution of the model and demonstrate that computing any correlation functions is tractable.
	We develop a coherent state approach, detailed in Appendix~\ref{appdix:coherent}, which enables us to derive an exact expression for the averaged grand partition function and provides a simple proof of the self-averaging properties of $\text{$R$-PSYK}_2$. Furthermore, this coherent state approach accurately captures the early-time behavior of the SFF. The exponential ramp behavior is investigated using the cluster function method, revealing a sharp transition in $R$-para-particle systems. The ramp exponent and the onset time $t_p$ are governed by a constant $C_0$. For the $R$-$\mathrm{PSYK}_2$ models listed in Table~\ref{table:six-cases}, the asymptotic behavior of $C_0$ depends on the parameter $m$: as $N \to \infty$, $C_0$ either approaches a finite constant or diverges ($C_0 \to \infty$), resulting in a transition in the SFF shape.  
	
	As an example, consider $z_R(x) = 1 + m x$. In this case, its SFF is
	\begin{equation}
		\begin{aligned}
			\mathcal{K}_{t\ll1}&=\frac{g_{0}^{N}e^{N\frac{r_{1}J_{1}(2t)}{t}\cos\left(\mu t\right)}}{D^{2N}}\exp\left[2N\left(B_{1}-\rho_{1}\right)\cos\left(\mu t\right)+N\left(B_{0}+\log\left(\rho_{1}^{2}+2\rho_{1}\cos(\mu t)+1\right)\right)\right]\co\\ \mathcal{K}_{1\ll t\ll N}&=\frac{g_{0}^{N}e^{N\frac{r_{1}J_{1}(2t)}{k}\cos\left(\mu t\right)}}{D^{2N}}\exp\left[NB_{0}+C_{0}t\right]\co\\ \mathcal{K}_{t\gg N}&=\frac{g_{0}^{N}e^{N\frac{r_{1}J_{1}(2t)}{t}\cos\left(\mu t\right)}}{D^{2N}}\co
		\end{aligned}
	\end{equation}
	where $J_1$ is the Bessel function. The constants $g_0$, $r_1$, $B_0$, and $C_0$ are functions of $m$, as defined in the main text. As discussed in \cite{Liao_2020}, $\text{SYK}_2$ exhibits an exponential ramp. Explicitly, we have
	\begin{align}
		\mathcal{K}_{t_p \ll t \ll N} \sim \exp(C_0 t)\ed
	\end{align}
	The transition time $t_p$ and exponent $C_0$ depend on $m$ as
	\begin{align}
		t_p \sim \begin{cases}
			N^{2/5} & m > 1, \\
			\left(\dfrac{N}{\ln N}\right)^{2/5} & m = 1,
		\end{cases}
		\quad
		C_0 \sim \begin{cases}
			\mathcal{O}(1), & m > 1, \\
			\mathcal{O}(\ln N), & m = 1.
		\end{cases}
	\end{align}
	This shows a dramatic transition between the fermionic SYK$_2$  (or trivial $R$-PSYK$_2$ with $m=1$)  and the nontrivial fermionic $R$-PSYK$_2$ ($m>1$). Moreover, the plateau height depends on $m$ via
	\begin{align}
		\frac{g_0^{N}}{D^{2N}} = \frac{(m^2 + 1)^N}{(m + 1)^{2N}} \ge \frac{1}{D^N} = \frac{1}{\text{dim}(\mathcal{H})},
	\end{align}
	indicating that the plateau for $R$-para-particles is larger than that of a regular chaotic system, as shown in Fig.~\ref{fig:SFF}.

	\paragraph{Structure of the paper}
	We study the thermodynamics in Section~\ref{sec:thermodynamics}. Using the coherent state approach, we derive analytical expressions for the ensemble-averaged partition function and prove the self-averaging property of the model. We also discuss the high-temperature expansion and compare the results for different cases of $R$-para-particle statistics.
	In Section~\ref{sec:SFF}, we study the two-point SFF of the $\text{$R$-PSYK}_2$ model in the large $N$ limit. Using both the coherent state approach and the cluster function method, we derive the SFF across its characteristic regimes: the early-time decay, the ramp, and the plateau. Our analysis reveals a sharp transition in the ramp dynamics, governed by the exponent $C_0$.
	In Section~\ref{sec:time-evolution}, we discuss the time evolution of operators in the model. We outline the methodology for calculating correlation functions, including both two-point functions and  OTOCs, while deferring explicit calculations to future work.
	In Section~\ref{sec:conclusion}, we summarize our findings, with particular emphasis on the transition in the SFF ramp behavior. We also suggest promising directions for future research, including studies of $\text{$R$-PSYK}_q$ models with $q>2$ and more detailed investigations of correlation functions.
	Finally, Appendix~\ref{appendix:cluster} provides detailed derivations of the cluster function method, while Appendix~\ref{appdix:coherent} presents the coherent state approach for evaluating ensemble-averaged quantities.

	\section{Thermodynamics}
	In this section, we begin by reviewing the Hilbert space construction of the model, following the approach outlined in \cite{Wang_2025}. This construction naturally leads to a factorized expression for the partition function $Z_{\beta,\mu}$. Using the coherent state approach, we then evaluate its ensemble average and derive an analytical expression in terms of Bessel functions. 
	We verify these results through two independent methods: high-temperature expansion and numerical simulations, finding excellent agreement between them. Furthermore, we examine a cluster approach implementation, which unfortunately fails to produce correct results. This leads us to conclude that the cluster method with a rough kernel is unsuitable for finite temperature scenarios. 
	\label{sec:thermodynamics} 
	
	\subsection{Constructing the Hilbert space}	
	A general Hamiltonian with quadratic interactions can be expressed as
	\begin{align}  
		H = \sum_{ijab} \left( h_{iajb}^{+-} \psi_{i,a}^+ \psi_{j,b}^- + h_{iajb}^{++} \psi_{i,a}^+ \psi_{j,b}^+ \right) + \text{h.c.},  
	\end{align}
	where $\text{h.c.}$ denotes the Hermitian conjugate. For simplicity, we focus on the case
	\begin{align}  
		H = \sum_{a=1}^{m} \sum_{i,j} (h_{ij} - \mu \delta_{ij}) \psi_{i,a}^+ \psi_{j,a}^-,  
	\end{align}
	where $h_{ij} = h_{ji}^*$ is drawn from the GUE. In the absence of disorder, this reduces to the model studied in \cite{Wang_2025}. Like the standard SYK model, it features all-to-all interactions.
	By applying a unitary transformation that diagonalizes $h_{ij}$,
	\begin{align}  
		\psi_{i,a}^- \to U_{ij} \psi_{j,a}^-, \quad \psi_{i,a}^+ \to  \psi_{j,a}^+ U_{ji}^\dagger,  
	\end{align}
	the new operators $\psi^{\pm}$ retain the same commutation relations, the Hamiltonian then simplifies to
	\begin{align}  
		H = \sum_{i=1}^N (\varepsilon_i - \mu) n_i, \quad n_i = \sum_{a=1}^m \psi_{i,a}^+ \psi_{i,a}^-,  
	\end{align}
	where $[n_i, n_j] = 0$ for $i \neq j$.  
	The vacuum state $\ket{0}$ is defined by
	\begin{align}  
		\psi_{i,a}^- \ket{0} = 0 \quad \forall \, i, a,  
	\end{align}
	and the $n$-particle states at each site are given by
	\begin{align}  
		\psi_{i,a_1}^+ \psi_{i,a_2}^+ \ldots \psi_{i,a_n}^+ \ket{0}.  
	\end{align}
	The construction of the full basis for this model is discussed in detail in \cite{Wang_2025}. For completeness, we briefly review the formulation of $n$-particle wave functions here.

	Let $\{\Psi^\alpha_{a_1a_2\ldots a_n}\}_{\alpha=1}^{d_n}$ be a complete set of linearly independent solutions to the system of linear equations
	\begin{equation}\label{eq:V_n_basis}
		\sum_{a'_j,a'_{j+1}} R^{a_ja_{j+1}}_{a'_ja'_{j+1}} \Psi_{a_1\ldots a'_ja'_{j+1}\ldots a_n} = \Psi_{a_1\ldots a_ja_{j+1}\ldots a_n}
	\end{equation}
	for $j = 1,2,\ldots,n-1$. For fermionic or bosonic systems, this reduces to totally antisymmetric or symmetric wavefunctions respectively. Given the unitarity of the $R$-matrix, we normalize the coefficients such that
	\begin{equation}\label{eq:singlemodewf_normalization}
		\sum_{a_1,\ldots,a_n} \Psi^{\beta*}_{a_1\ldots a_n} \Psi^\alpha_{a_1\ldots a_n} = \delta_{\alpha\beta}.
	\end{equation}
	These $n$-particle eigenfunctions can be used to construct a basis for our model. We then define the creation operator~\footnote{One can introduce \textbf{local} spin operators $\hat{y}_{i,a}^\pm$ that satisfy the same exchange relations as $\psi_{i,a}^\pm$, but commute between different sites. By a similar method, we construct \textbf{local} Hilbert space for each site, then one can see $\psi_{i,a}^{\pm}$ is a global operator represented by MPOs, which leads to a simple proof for Eq. \eref{eq:full_basis_orthonormal}.  }
	\begin{equation}\label{eq:single_mode_creation}
		\hat{\Psi}^{(i)+}_{n,\alpha} \equiv \frac{1}{\sqrt{n!}} \sum_{a_1\ldots a_n} \Psi^\alpha_{a_1\ldots a_n} \hat{\psi}^+_{i,a_1} \cdots \hat{\psi}^+_{i,a_n}.
	\end{equation}
	The complete orthonormal basis for the state space consists of states of the form
	\begin{equation}\label{eq:full_basis}
		\ket{{}^{\alpha_1}_{n_1}, {}_{n_2}^{\alpha_2}, \ldots, {}_{n_N}^{\alpha_N}} 
		= \hat{\Psi}^{(1)+}_{n_1,\alpha_1} \hat{\Psi}^{(2)+}_{n_2,\alpha_2} \cdots \hat{\Psi}^{(N)+}_{n_N,\alpha_N} \ket{0},
	\end{equation}
	where the quantum numbers $\{(n_i,\alpha_i)\}_{i=1}^N$ (with $1 \leq \alpha_i \leq d_{n_i}$) can be chosen independently. Then one can prove the orthonormality
	\begin{equation}\label{eq:full_basis_orthonormal}
		\braket{{}^{\beta_1}_{n'_1}, {}_{n'_2}^{\beta_2}, \ldots, {}_{n'_N}^{\beta_N} | {}^{\alpha_1}_{n_1}, {}_{n_2}^{\alpha_2}, \ldots, {}_{n_N}^{\alpha_N}} 
		= \prod_{j=1}^N \delta_{n_jn'_j} \delta_{\alpha_j\beta_j}.
	\end{equation}
	For $R^{ab}_{cd} = -\delta_{ac}\delta_{bd}$, solving Eq.~\eqref{eq:V_n_basis} yields $d_0=1$, $d_1=m$, and $d_{n\geq 2}=0$. Similarly, for $R^{ab}_{cd} = \lambda_{ab}\xi_{cd} - \delta_{ac}\delta_{bd}$, we obtain $d_0=1$, $d_1=m$, $d_2=1$, and $d_{n>2}=0$. These two cases serve as our primary examples throughout this work. 
	The methodology presented here naturally extends to any fermionic $R$-PSYK$_2$ model with a generating polynomial $z_{R}(x) = \sum_{n=0}^L d_n x^n$ (where $d_{n>L} = 0$) and its corresponding bosonic dual.

	For the basis states in Eq.~\eqref{eq:full_basis}, the energy is given by $E = \sum_{k=1}^N n_k(\varepsilon_k - \mu)$, which leads the partition function
	\begin{align}
		Z_\beta = \prod_{i=1}^{N} \sum_{n_{i}=0}^{L} d_{n_{i}} e^{-\beta(\varepsilon_{i}-\mu)n_{i}}\ed
	\end{align}
	Our primary interest lies in ensemble-averaged observables $\mathbb{E}(A)$, where $A$ represents thermodynamic quantities such as energy and entropy. Importantly, these must be calculated from $\mathbb{E}(\log Z_\beta)$ rather than $\log \mathbb{E}(Z_\beta)$. 
	For the regular SYK model, $\mathbb{E}(Z_\beta)$ can be conveniently computed using saddle-point analysis in the large-$N$ limit through path integrals. The SYK model exhibits self-averaging behavior, satisfying $\log \mathbb{E}(Z_\beta) = \mathbb{E}(\log Z_\beta)$ for large $N$. 
	The situation differs markedly for the $R$-PSYK$_2$ model: since $Z_\beta$ factorizes into independent site contributions, $\mathbb{E}(\log Z_\beta)$ becomes significantly easier to compute. In this section, we employ the coherent state approach to evaluate $\mathbb{E}(Z_\beta)$. Remarkably, we prove that the $R$-PSYK$_2$ model remains exactly self-averaged even in the $R$-para-boson case.

	\subsection{Coherent state approach}
	As the partition function of $R$-PSYK$_2$ factorized for each mode, we can deal with it using the coherent state approach derived in Appendix \ref{appdix:coherent}. Here we have
	\begin{align}
		\mathcal{G}\left(2u\cos\theta\right)=\sum_{n=0}^{L}d_{n}e^{-n\beta(2u\cos\theta-\mu)}\ed
	\end{align}
	Then we consider the expression for large $N$
	\begin{align}
		\frac{1}{N}\log \mathbb{E}(Z_\beta) &=
		\int_{0}^{1}\int_{0}^{2\pi}\frac{udud\theta}{\pi}\log\left[\widetilde{D}\left(1+f_{\beta}\left(u\cos\theta\right)\right)\right]
	\end{align}
	where we have defined 
	\begin{align}\label{eq:definitions}
		\widetilde{D}&=\sum_{n=0}^{L}{d}_{n}e^{n\beta \mu},~\widetilde{d}_{n,\beta\mu}\equiv {d_{n}e^{n\beta\mu}\over \widetilde{D}}, \nn
		f_{\beta}\left(u\cos\theta\right)&\equiv -1+\sum_{n=0}^{L}\widetilde{d}_{n,\beta\mu}e^{-2n\beta u\cos\theta}=-1+\frac{1}{\widetilde{D}}z_{R}\left(e^{-\beta\left(2u\cos\theta-\mu\right)}\right)\ed
	\end{align}
	Notice that $|f_{\beta}|\le1,\text{for}\ \cos\theta\not=0$, so we can safely expand the  logarithmic function 
	\begin{align}\label{eq:Z_log-exp}
		\frac{1}{N}\log \mathbb{E}(Z_\beta)= \log\widetilde{D}+\int_{0}^{1}\int_{0}^{2\pi}\frac{udud\theta}{\pi}\sum_{j=1}^{\infty}\frac{(-1)^{j-1}}{j}f_{\beta}^{j}\ed
	\end{align}
	Using the integral formula
	\begin{align}
		\int_{0}^{1}\int_{0}^{2\pi}\frac{udud\theta}{\pi}\exp\left[2ru\cos\theta\right]=\frac{2I_{1}(|2r|)}{|2r|}\co
	\end{align}
	we can always write ${1\over N}\log \mathbb{E}(Z_\beta)$ via $I_n(x)$ the modified Bessel functions  of the first kind.

	We then test if the order of taking ensemble average matters i.e., $\mathbb{E}\left(\log Z_{\beta}\right)\stackrel{?}{=}\log\mathbb{E}\left(Z_{\beta}\right)$. 
	For any quantity that has the form $\log Z_\beta=\sum_{i=1}^{N}\mathcal{G}\left(\varepsilon_{i}\right)$,  we have 
	\begin{align}
		\mathbb{E}\left(\log Z_\beta\right)=N\int d\varepsilon\rho(\varepsilon)\mathcal{G}\left(\varepsilon\right)\ed
	\end{align}
	Here 
	\begin{align}
		\mathcal{G}\left(\varepsilon\right)=\log\left[\sum_{n=0}^{\infty}d_{n}e^{-\beta(\varepsilon-\mu)n}\right]=\log\widetilde{D}+\log\left[1+f_{\beta}\right],
	\end{align}
	where $f_{\beta}(\varepsilon)\equiv -1+\sum_{n=0}^{\infty}\widetilde{d}_{n}e^{-n\beta\varepsilon}$. Using $\int d\varepsilon\rho(\varepsilon)e^{-r\varepsilon}=\,_{0}F_{1}\left(2;r^{2}\right)=\frac{2I_{1}(|2r|)}{|2r|}$, we exactly have  
	\begin{align}
		\mathbb{E}\left(\log Z_{\beta}\right)=\log\mathbb{E}\left(Z_{\beta}\right)
	\end{align}
	as expected for SYK model. The proof relies on the validity of the expansion of the logarithm function, which always holds for finite $D$, all kinds of $R$-para-fermions.  For a general single model partition function, we have 
	\begin{align}
		\int d\varepsilon\rho(\varepsilon)\log\left(z_{R}\left(e^{-\beta\left(\varepsilon-\mu\right)}\right)\right)\stackrel{?}{=}\int_{0}^{1}\int_{0}^{2\pi}\frac{udud\theta}{\pi}\log\left(z_{R}\left(e^{-\beta\left(2u\cos\theta-\mu\right)}\right)\right)\ed
	\end{align} 
	If both sides have the same expansion series of $x=e^{-\beta\left(\varepsilon-\mu\right)}$ (or $x=e^{-\beta\left(2u\cos\theta-\mu\right)}$) at the origin $x=0$ in the integration, we then have $\mathbb{E}\left(\log Z_{\beta}\right)=\log\mathbb{E}\left(Z_{\beta}\right)$. Notice $\varepsilon \in [-2,2]$ $u\in[0,1]$, and $\theta \in [0,2\pi)$, $x$ has the same variation region on both integrals. So we just need to evaluate the validity of the expansion of $ f(x)=\log z_R(x)$ at the origin of the integration. It is obvious that we need to require $|x|$ is smaller than a radius of  convergence $\mathtt{R}$, explicitly
	\begin{align}\label{eq:self_average_condition}
		e^{-\beta(\varepsilon-\mu)}<\mathtt{R},\forall \varepsilon \in [-2,2] \Rightarrow e^{\beta(\mu+2)}< \mathtt{R}
	\end{align}     
	where $\mathtt{R}$ can be obtained by evaluating equations 
	\begin{align}
		z_R(x)=\infty\ \text{or}\ z_R(x)=0	
	\end{align} 
	which determines the singularities of integrand $\log(z_R(x))$.   For $R$-para-fermions, $z_R(x)$ is a polynomial of $x$ with positive coefficients $d_n$, so that $\mathtt{R}=\infty$, the $R$-PSYK$_2$ is always self-averaged. For $R$-para-bosons, there are always singularities for $z_R(x)$ is a rational function, so that $\mathtt{R}=\text{finite}$, we need to impose Eq. \eref{eq:self_average_condition}.
	Taking ordinary boson as an example, we have $z_R(x)=(1-x)^{-m}$ so that $\mathtt{R}=1$, which implies the bosonic SYK$_2$ is self-averaged for $\mu<-2$. For the dual $R$-para-bosons of the $R$-para-fermion with $z_{R}(x)=\sum_{n=0}^{L} d_n x^n$, we have 
	\begin{align}
		z_{\text{ParaB}}(x)={1\over \sum_{n=0}^{L} (-1)^n d_n x^n}\ed
	\end{align} 
	Then
	\begin{align}
		{\log \mathbb{E}(Z_{\text{ParaB};\beta})\over N}=-\int_{0}^{1}\int_{0}^{2\pi}\frac{udud\theta}{\pi}\log\left(\sum_{n=0}^{L} (-1)^n d_ne^{-n\beta\left(2u\cos\theta-\mu\right)}\right)\ed
	\end{align}
	One can use a similar method to deal with the integral. Actually, we have 
	\begin{align}
		{\log \mathbb{E}(Z_{\text{ParaB};\beta})\over N}=-{\log \mathbb{E}(Z_{\text{ParaF};\beta})\over N}\bigg\vert_{e^{\beta\mu}\to -e^{\beta\mu}}
	\end{align}
	where $\mathbb{E}(Z_{\text{ParaF};\beta})$ is the partition function of dual $R$-para-fermionic SYK$_2$ with $z_{R}(x)=\sum_{n=0}^{L} d_n x^n$. 
	Since we only consider the self-averaged theory, we denote the averaged partition function by $\mathcal{Z}_{\beta,\mu}=\mathbb{E}(Z_{\beta})$,  then we can calculate all other thermodynamic functions. 
	We will test the validity of the coherent state approach displayed here by calculating the three $R$-para-fermionic examples listed in Table \ref{table:six-cases} and compare their high-temperature expansion with the numerical simulation results. 
	\subsubsection{Ordinary Fermions and Bosons}
	As a warmup, we first consider ordinary fermions with $m$ flavors. Since $z_R=(1+x)^m$, we have 
	\begin{align}\label{eq:Z_fermion}
		{\log \mathcal{Z}_F\over N} =m \int_{0}^{1}\int_{0}^{2\pi}\frac{udud\theta}{\pi}\log\left(1+e^{-\beta\left(2u\cos\theta-\mu\right)}\right)=m\sum_{j=1}^{\infty}\frac{\left(-\right)^{j-1}e^{j\beta\mu}}{j}\frac{2I_{1}\left(2j\beta\right)}{2j\beta}\ed
	\end{align} 
	Then we consider the high-temperature expansion (where we do not expand $e^{j\beta\mu}$)
	\begin{align}
		{\log \mathcal{Z}_{F }\over mN}=&\log\left(e^{\beta\mu}+1\right)+\frac{\beta^{2}e^{\beta\mu}}{2\left(e^{\beta\mu}+1\right)^{2}}+\frac{\beta^{4}e^{\beta\mu}\left(-4e^{\beta\mu}+e^{2\beta\mu}+1\right)}{12\left(e^{\beta\mu}+1\right)^{4}}+\mathcal{O}(\beta^5)\ed
	\end{align}
	Then we deal with ordinary bosons. For $z_R(x)=(1-x)^{-m}$, we have 
	\begin{align}\label{eq:Z_boson}
		\frac{\log\mathcal{Z}_B}{N}=-m\int_{0}^{1}\int_{0}^{2\pi}\frac{udud\theta}{\pi}\log\left(1-e^{-\beta\left(2u\cos\theta-\mu\right)}\right)=m\sum_{j=1}^{\infty}\frac{e^{j\beta\mu}}{j}\frac{2I_{1}\left(2j\beta\right)}{2j\beta}
	\end{align}
	where we must impose $\mu<0$ to make the summation convergent. As discussed before, one can obtain the partition function for ordinary bosons simply by replacing $e^{\beta\mu}\to -e^{\beta\mu}$, one can check it by simply comparing Eq. \eref{eq:Z_boson} with Eq. \eref{eq:Z_fermion}.  Its high-temperature expansion is 
	\begin{align}
		\frac{1}{mN}\log\mathcal{Z}_B=&-\log\left(1-e^{\beta\mu}\right)+\frac{\beta^{2}e^{\beta\mu}}{2\left(e^{\beta\mu}-1\right)^{2}}+\frac{\beta^{4}e^{\beta\mu}\left(4e^{\beta\mu}+e^{2\beta\mu}+1\right)}{12\left(e^{\beta\mu}-1\right)^{4}}+\mathcal{O}(\beta^5)\ed
	\end{align}
	Although the formula makes sense for any $\mu <0$, it is derived by expanding the logarithm function in the integrand of Eq.~\eref{eq:Z_boson}, which is only valid for $\mu<-2$. Notice that $-2$ is the lower bound of semicircle density $\rho(\varepsilon)$ in the large $N$ limit. For large but finite $N$, we may have few energy levels $\varepsilon_i<\mu $. So for numerical simulation, we regularize the definition partition function for the bosonic SYK$_2$
	\begin{align}
		\mathbb{E}_{reg}(Z_\beta)=\text{Re}\left(\prod_{i=1}^{N}\left(\frac{1}{1+\delta\sqrt{-1}-e^{-\beta(\varepsilon_{i}-\mu)}}\right)^{m}\right)
	\end{align} 
	where $\delta\ll 1$ is a small number choice to make the numerical simulation stable. 
	
	\subsubsection{$R$-para-fermions with $z_R=1+mx$ }
	For Example A with $z_R=1+mx$, using Eq.~\eref{eq:definitions}, we have 
	\begin{align}
		f_{\beta}\left(u\cos\theta\right)=-1+\widetilde{d}_{0}+\widetilde{d}_{1}e^{-2\beta u\cos\theta}\ed
	\end{align}
	Here  $d_0=1,d_1=m$, it is easy to obtain the explicit expression of $\widetilde{d}_0,\widetilde{d}_1$, but we just keep the form. It is direct to obtain
	\begin{equation}
		\begin{aligned}
			{\log \mathcal{Z} \over N}=&\log\widetilde{D}+\sum_{j=1}^{\infty}\sum_{k=0}^{j}\frac{\left(-\right)^{j-1}}{j}\binom{j}{k}\left(\widetilde{d}_{1}\right)^{k}\left(\widetilde{d}_{0}-1\right)^{j-k}\frac{2I_{1}\left(2k\beta\right)}{2k\beta}\\
			%=&\log\widetilde{D}+\sum_{k=0}^{\infty}\sum_{j=k,j>0}^{\infty}\frac{\left(-\right)^{j-1}}{j}\binom{j}{k}\left(\widetilde{d}_{1}\right)^{k}\left(\widetilde{d}_{0}-1\right)^{j-k}\frac{2I_{1}\left(2k\beta\right)}{2k\beta}\\
			=&\log\widetilde{D}+\log\widetilde{d}_{0}+\sum_{k=1}^{\infty}\frac{(-1)^{k+1}}{k}\left(\frac{\widetilde{d}_{1}}{\widetilde{d}_{0}}\right)^{k}\frac{2I_{1}\left(2k\beta\right)}{2k\beta}
		\end{aligned}
	\end{equation}
	where we use the trick to exchange the summation for $j$ and $k$: $\sum_{j=1}^{\infty}\sum_{k=0}^{j}\to\sum_{k=0}^{\infty}\sum_{j=k,j>0}^{\infty}$. To test the validity of the formula, we consider its high temperature expansion
	\begin{align}
		{\log \mathcal{Z}\over N}=&\log\widetilde{D}+\log\widetilde{d}_{0}+\log(r+1)+\frac{\beta^{2}r}{2(r+1)^{2}}+\frac{\beta^{4}r\left(r^{2}-4r+1\right)}{12(r+1)^{4}}+\mathcal{O}(\beta^6)\ed
	\end{align}
	Here we have defined $r=\frac{\widetilde{d}_{1}}{\widetilde{d}_{0}}$ for simplicity. 
	\subsubsection{$R$-para-fermions with $z_R=1+mx+x^2$}
	For Example B with $z_R=1+mx+x^2$, using Eq.~\eref{eq:definitions}, we have
	\begin{align}
		f_{\beta}\left(u\cos\theta\right)=-1+\widetilde{d}_{0}+\widetilde{d}_{1}e^{-2\beta u\cos\theta}+\widetilde{d}_{2}e^{-4\beta u\cos\theta}\ed
	\end{align}
	Here $d_0=1,d_1=m,d_2=1$, it is easy to obtain the explicit expression of $\widetilde{d}_0,\widetilde{d}_1,\widetilde{d}_2$, we just keep the form, so that the result can easily apply to any $R$-PSYK$_2$ with $z_R(x)=d_0+d_1x+d_2 x^2$. A direct calculation leads to 
	\begin{align}
		{\log \mathcal{Z} \over N}=\log\widetilde{D}+\sum_{j=1}^{\infty}\sum_{k=0}^{j}\sum_{l=0}^{k}\frac{\left(-\right)^{j-1}}{j}\binom{j}{k}\binom{k}{l}\left(\widetilde{d}_{0}-1\right)^{j-k}\left(\widetilde{d}_{2}\right)^{l}\left(\widetilde{d}_{1}\right)^{k-l}\frac{2I_{1}\left(2(k+l)\beta\right)}{2(k+l)\beta}\ed
	\end{align}
	The high-temperature expansion yields
	\begin{equation}
		\begin{aligned}
			{\log \mathcal{Z} \over N}=&\log\widetilde{D}+\log\left(\widetilde{d}_{0}+\widetilde{d}_{1}+\widetilde{d}_{2}\right)+\frac{\beta^{2}\left(\widetilde{d}_{1}\widetilde{d}_{2}+\widetilde{d}_{0}\left(\widetilde{d}_{1}+4\widetilde{d}_{2}\right)\right)}{2\left(\widetilde{d}_{0}+\widetilde{d}_{1}+\widetilde{d}_{2}\right)^{2}}\\
			&+\frac{\beta^{4}}{12\left(\widetilde{d}_{0}+\widetilde{d}_{1}+\widetilde{d}_{2}\right)^{4}}
			\Bigg[\left(\widetilde{d}_{1}+16\widetilde{d}_{2}\right)\widetilde{d}_{0}^{3}-\left(4\widetilde{d}_{1}^{2}+13\widetilde{d}_{2}\widetilde{d}_{1}+64\widetilde{d}_{2}^{2}\right)\widetilde{d}_{0}^{2}\\
			&+\left(\widetilde{d}_{1}^{3}+8\widetilde{d}_{2}\widetilde{d}_{1}^{2}-13\widetilde{d}_{2}^{2}\widetilde{d}_{1}+16\widetilde{d}_{2}^{3}\right)\widetilde{d}_{0}+\widetilde{d}_{1}\widetilde{d}_{2}\left(\widetilde{d}_{1}^{2}-4\widetilde{d}_{2}\widetilde{d}_{1}+\widetilde{d}_{2}^{2}\right)\Bigg]+\mathcal{O}(\beta^5)\ed
		\end{aligned}
	\end{equation}

	\subsection{Cluster function approach}
\begin{figure}[h]
		\begin{center}
			\includegraphics[width=0.4\textwidth]{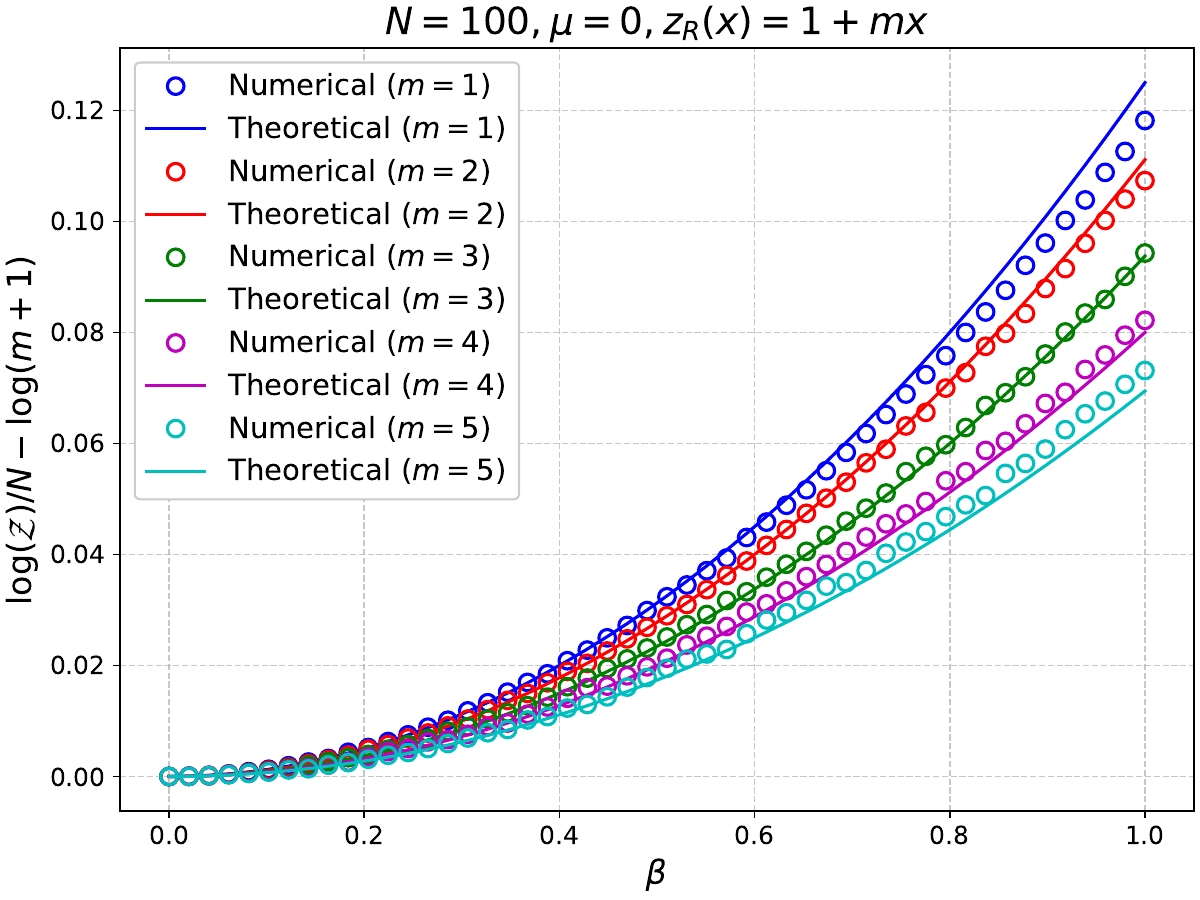}
			\includegraphics[width=0.4\textwidth]{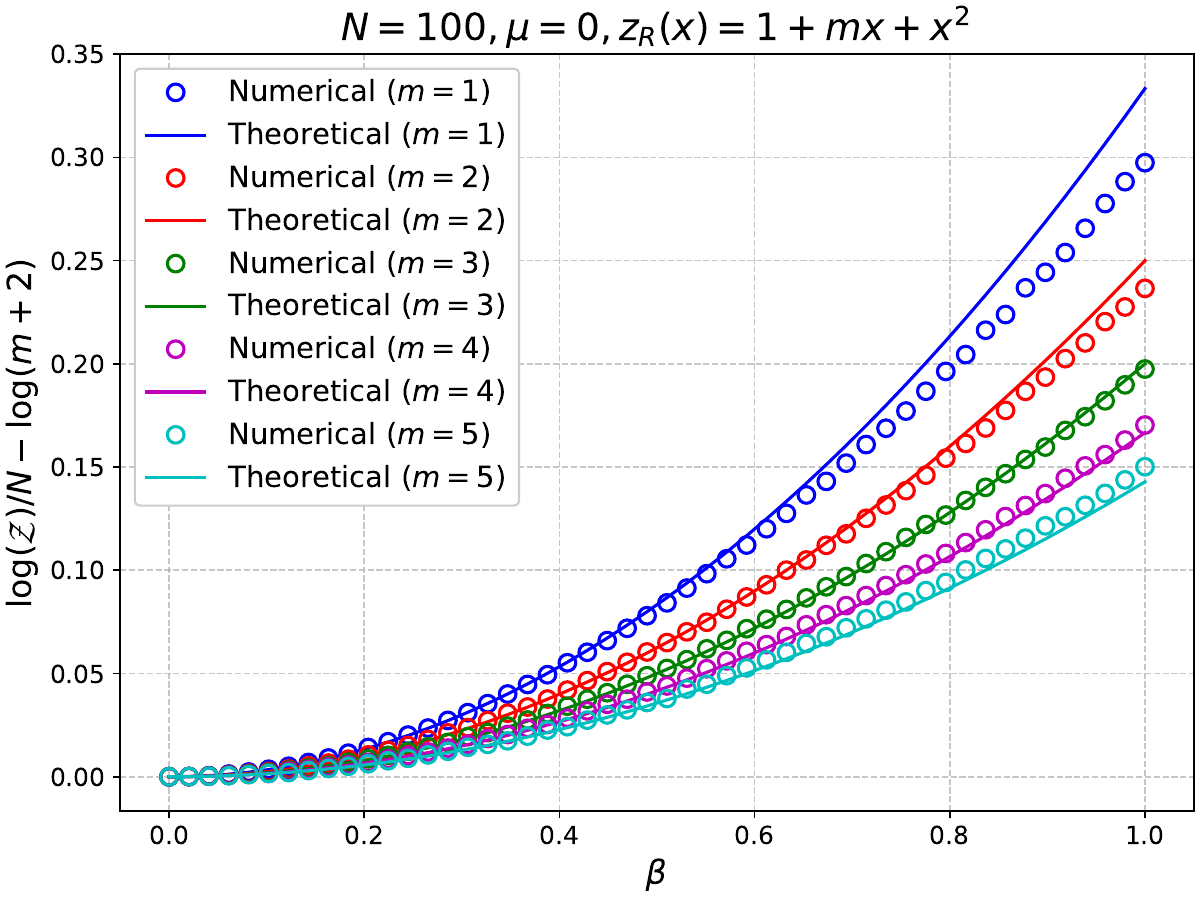}
			\caption{The plot compares the ensemble-averaged partition function for numerical simulations and theoretical expectations in two cases. For $z_R(x) = 1 + m x$, the theoretical expectation is $\log\mathcal{Z}/N = \log(1+m) + \frac{m}{2(1+m)^2}\beta^2 + \mathcal{O}(\beta^4)$. For $z_R(x) = 1 + m x + x^2$, it becomes $\log\mathcal{Z}/N = \log(1+m) + \frac{1}{2+m}\beta^2 + \mathcal{O}(\beta^4)$. The numerical results are averaged over 200 samples. }
			\label{fig:Z-highT}
		\end{center}
	\end{figure}

	The cluster function approach is used to calculate the ensemble averaged SFF in \cite{Liao_2020}. As derived in Appendix \ref{appendix:cluster}, we expected the cluster function approach to still work for high-temperature cases. 
	For simplicity, we denote 
	\begin{align}
		\sum'_{\{k\}}\equiv \sum_{\{k\},k_{i}=1}^{L},~|\{k\}|\equiv \sum_{i=1}^n k_i \ed
	\end{align}
	Finally we have 
	\begin{align}
		\mathbb{E}(Z_{\beta})=\exp\left[N\sum_{n=1}^{N}\frac{\left(-1\right)^{n-1}}{n}\sum'_{\{k\}}d_{\{k\}}e^{\sum_{i=1}^{n}\beta k_{i}\mu}I_{n}^{E}(\beta\{k\})\right]\ed
	\end{align}
	With the explicit expressions of $I_{n}^{E}(\beta\{k\})$, we have 
	\begin{align}\label{eq:Z_highT}
		\frac{\log\mathbb{E}(Z_{\beta})}{N}&=\sum'_{k}d_{k}e^{\beta k\mu}\ _{0}F_{1}\left(2;k^{2}\beta^{2}\right)+\sum_{n=2}^{N}\frac{\left(-1\right)^{n-1}}{n}\sum'_{\{k\}}d_{\{k\}}e^{\sum_{i=1}^{n}\beta k_{i}\mu}\frac{\sinh\left(\frac{\pi\beta}{2}\sum_{i=1}^{n}k_{i}\right)}{\frac{\pi\beta}{2}\sum_{i=1}^{n}k_{i}}\co
	\end{align}
	where $\ _{0}F_{1}(x)$ is the confluent hypergeometric function. We replace the summation $\sum_{n=2}^N$ with $\sum_{n=2}^\infty$ to obtain the correct result for $\beta=0$. This leads to alternating series that are not absolutely convergent. For example, when $z_R(x)=1+mx$ and $\mu=0$, we have
	\begin{align}
		\frac{\log \mathbb{E}(Z_{\beta=0})}{N}=m+\sum_{n=2}^{\infty}\frac{\left(-1\right)^{n-1}}{n}m^n\to \log(1+m)\ed
	\end{align} 
	Therefore, we must interpret the summation in the definition of $\log Z_\beta$ as a formal expansion of $\log(1+m)$, which is strictly invalid for $m \geq 1$. To evaluate Eq.~\eref{eq:Z_highT}, we recommend first performing the high-temperature expansion and then computing the summation.
	The high-temperature expansion can be derived by expanding for $\beta \ll 1$
	\begin{align}
		\,_{0}F_{1}\left(2;\zeta^{2}\right)= 1+\frac{\zeta^{2}}{2}+\mathcal{O}(\zeta^4),~
		\frac{\sinh\left(x\right)}{x}= 1+\frac{x^{2}}{6}+\mathcal{O}\left(x^{4}\right)\ed
	\end{align}
	For example, for the case $z_R(x)=1+mx$, and $\mu =0$, we have 
	\begin{align}\label{eq:Z_exp}
		{\log \mathcal{Z}_{\beta\ll 1}\over N}=\log(m+1)+\frac{m}{24}\left(12-\frac{\pi^{2}m(m+2)}{(m+1)^{2}}\right)\beta^{2}+\mathcal{O}(\beta^4)\co
	\end{align}
	while the coherent state approach gives 
	\begin{align}\label{eq:Z-exp-compare}
		\frac{\log \mathcal{Z}_{\beta\ll 1}}{N}=&\log\left(m+1\right)+\frac{\beta^{2}m}{2(m+1)^{2}}+\mathcal{O}(\beta^4)\ed
	\end{align}
	Since numerical simulations have confirmed the validity of the coherent state approach (as demonstrated in Fig.~\ref{fig:Z-highT}), we observe that the cluster function obtained from the rough kernel method is unreliable. Instead, one could consider using the refined kernel
	\begin{align}  
		\widetilde{K}(\varepsilon_{i},\varepsilon_{j}) = \frac{\sin\left(N\pi(\varepsilon_{i} - \varepsilon_{j}) \rho\left(\frac{\varepsilon_{i} + \varepsilon_{j}}{2}\right)\right)}{\pi(\varepsilon_{i} - \varepsilon_{j})},
	\end{align}  
	though this would complicate analytical evaluation. Given that the coherent state approach has yielded satisfactory results, we will not pursue the refined kernel method further.
	\section{SFF}
	\label{sec:SFF}
	In this section, we evaluate the two-point SFF for the fermionic $R$-PSYK$_2$ model in the large $N$ limit. The SFF is defined as
	\begin{align}
		K(t) = \frac{1}{D^{2N}} \prod_{i=1}^{N} \abs{\sum_{n_{i}=0}^{L} d(n_{i}) e^{i t (\varepsilon_{i} - \mu) n_{i}}}^2,
	\end{align}
	where the prefactor $\frac{1}{D^{2N}}$ ensures the normalization condition $K(0) = 1$. We are particularly interested in its disorder-averaged value:
	\begin{align}
		\mathcal{K}(t) \equiv \mathbb{E}\left[K(t)\right].
	\end{align}
	As mentioned in the introduction, the SFF serves as a diagnostic tool for quantum chaos. Universally, it exhibits a rapid decay at early times ($t \ll 1$) and saturates to a plateau at late times ($t \gg N$). For chaotic quantum systems, the SFF typically displays a linear ramp $\mathcal{K}(t) \sim t$ in the intermediate region $t_p \ll t \ll N$. However, as shown in \cite{Liao_2020,Winer_2020}, the SYK$_2$ model exhibits an exponential ramp $\mathcal{K}(t) \sim \exp(C_0 t)$. The onset timescale $t_p$ and the exponent $C_0$ of this exponential ramp can be determined analytically.
	
	For the $R$-PSYK$_2$ model, we employ methods from random matrix theory to compute the SFF. In addition to the cluster function approach presented in \cite{Liao_2020}, we develop a coherent state approach (detailed in Appendix \ref{appdix:coherent}) that is particularly effective for times $t \sim \mathcal{O}(1)$ in the large $N$ limit. Numerical results confirm that the coherent state approach accurately captures the early-time behavior ($t \sim \mathcal{O}(1)$), whereas the cluster function approach loses precision in this region due to the roughness of the box approximation.
	
	For the intermediate time region, we rely on the cluster function approach and provide a general analysis for $R$-PSYK$_2$ with arbitrary polynomial weight functions $z_R(x) = \sum_{k=0}^L d_k x^k$. Our findings reveal a striking difference in ramp behavior: for $z_R(x) = 1 + m x$ ($m \not= 1$) and $z_R(x) = 1 + m x + x^2$ ($m \neq 2$), the exponent $C_0$ approaches a constant, whereas for the conventional SYK$_2$ model ($z_R(x) = 1 + x$ or $z_R(x) = (1 + x)^2$), $C_0$ scales as $\log N$. This leads to a dramatic change in the ramp behavior.

	\subsection{Coherent state approach}
	Before proceeding with calculations, we establish some key notations. We define
	\begin{equation}\label{eq:zRsq}
		|z_R|^2(t) \equiv \abs{\sum_{n=0}^{L}d_ne^{it(\varepsilon-\mu)n}}^2 \equiv \sum_{k=0}^L g_k\cos(kt(\varepsilon-\mu))\co
	\end{equation}
	with normalized coefficients $\widetilde{g}_k = g_k/D^2$ satisfying $\sum_{k=0}^L \widetilde{g}_k = 1$.
	\begin{figure}[h]
		\centering
		\includegraphics[width=0.4\textwidth]{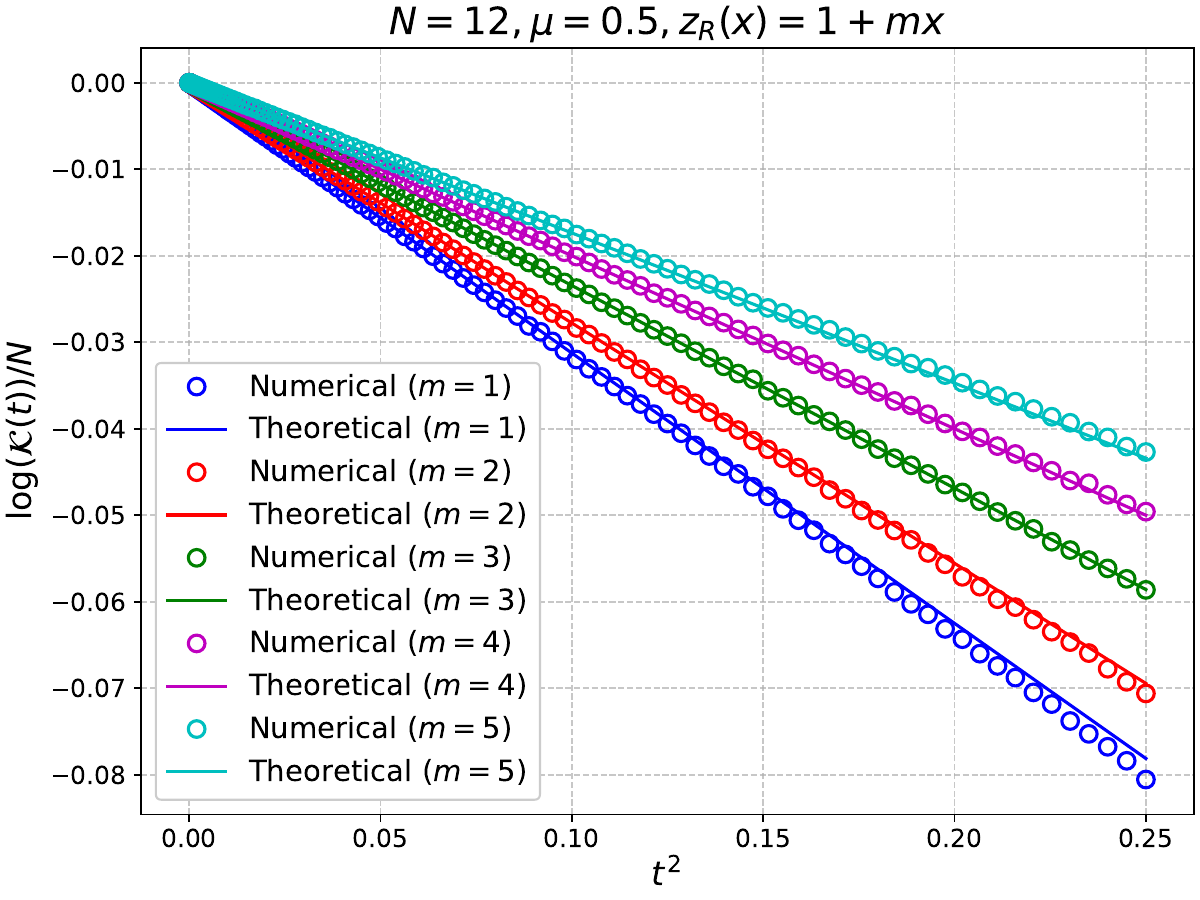}
		\includegraphics[width=0.4\textwidth]{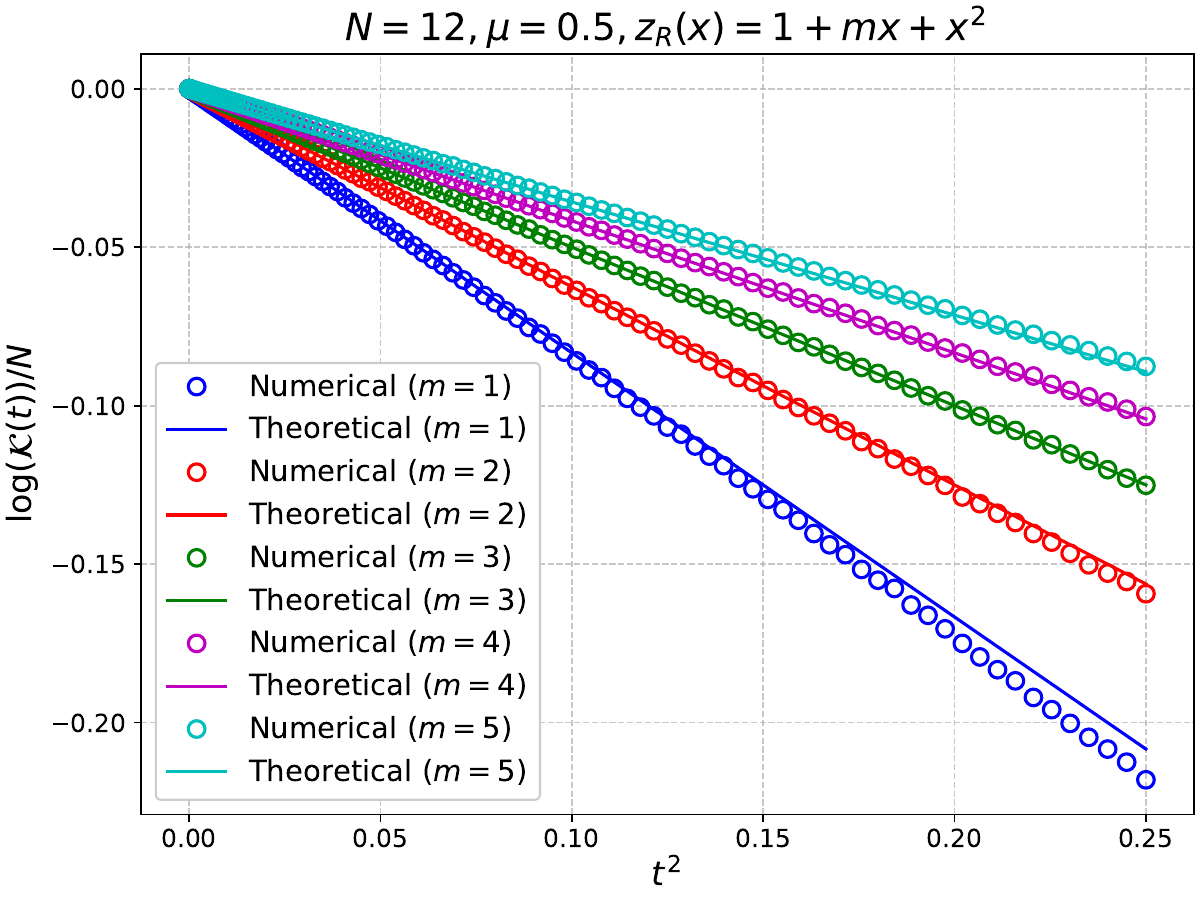}
		\caption{Short-time SFF comparison between numerical simulations (averaged over 2000 samples) and theoretical predictions from the coherent state approach, for $z_R(x)=1+mx$ (left) and $z_R(x)=1+mx+x^2$ (right).}
		\label{fig:SFF-short-time}
	\end{figure}
	The ensemble-averaged SFF can be evaluated using the techniques developed in Appendix \ref{appdix:coherent}. However, two approximations limit the validity of this approach: (1) changing the order of taking the cut-off and determinant in Eq.~\eqref{eq:G_trace}, and (2) replacing the cut-off trace $\text{tr}_N$ with the coherent state basis trace $\text{tr}_R$ in Eq.~\eqref{eq:trN-to-trR}. Consequently, this method is only reliable for $t \sim \mathcal{O}(1)$ in the large $N$ limit and cannot capture the ramp behavior of $R$-PSYK$_2$.
	
	While terms proportional to $1/R = 1/\sqrt{N}$ could be neglected, we choose to retain them. Applying the Baker-Campbell-Hausdorff formula
	\begin{equation}
		e^{X}e^{Y} = e^{Z}, \quad Z = X + Y + \frac{1}{2}[X,Y] + \cdots \co
	\end{equation}
	we derive the coherent state representation
	\begin{equation}
		e^{a+a^\dagger} = e^{a^\dagger}e^{a}e^{1/2} = e^{\alpha+\alpha^{*}-1/2}\co
	\end{equation}
	where we have used the substitutions $a^\dagger \to \alpha^{*}-\partial_{\alpha}$ and $a \to \alpha$ in the coherent basis. As derived in Appendix \ref{appdix:coherent}, we make the replacement $\varepsilon \to 2u\cos\theta - \frac{1}{2R}$ in Eq.~\eqref{eq:zRsq}, yielding
	\begin{equation}
		\mathcal{G}(2u\cos\theta) = \sum_{k=0}^{L} g_k \cos\left(kt\left(2u\cos\theta - \mu_R\right)\right),
	\end{equation}
	where we have defined the shifted chemical potential $\mu_R \equiv \mu + \frac{1}{2R} = \mu + \frac{1}{2\sqrt{N}}$. 
	The ensemble-averaged SFF then takes the form
	\begin{equation}
		\frac{\log\mathcal{K}(t)}{N} = -2\log D + \int_{0}^{1}\int_{0}^{2\pi} \frac{u\,du\,d\theta}{\pi} \log\left[\sum_{k=0}^{L} g_k \cos\left(kt\left(2u\cos\theta - \tfrac{1}{2R} - \mu\right)\right)\right].
	\end{equation}
	In the formal large-$t$ limit ($t\to\infty$), where we can ignore oscillatory terms, this simplifies to the plateau value
	\begin{equation}\label{eq:coherence-plateau}
		\frac{\log \mathcal{K}_{t=\infty}}{N} = -2\log D + \log g_0\co
	\end{equation}
	which correctly reproduces the expected plateau behavior.
	For $t\ll 1$, we have the following expansion 
	\begin{equation}\label{eq:coherent-SFF-early}
		\begin{aligned}
			\frac{\log\mathcal{K}(t)}{N}&=\int_{0}^{1}\int_{0}^{2\pi}\frac{udud\theta}{\pi}\log\left[1-\frac{1}{2}\sum_{k=0}^{L}\widetilde{g}_{k}k^{2}t^{2}\left(2u\cos\theta-\mu_R\right)^{2}\right]+\mathcal{O}(t^{4})\\&=-\frac{1}{2}\sum_{k=0}^{L}\widetilde{g}_{k}k^{2}t^{2}\int_{0}^{1}\int_{0}^{2\pi}\frac{udud\theta}{\pi}\left(2u\cos\theta-\mu_R\right)^{2}+\mathcal{O}(t^{4})\\&=-\frac{1+\mu_{R}^{2}}{2}\sum_{k=0}^{L}\widetilde{g}_{k}k^{2}t^{2}+\mathcal{O}(t^{4})\ed
		\end{aligned}
	\end{equation}
	Numerical verification of these results is shown in Fig.~\ref{fig:SFF-short-time}. For the time scale $t \sim \mathcal{O}(1)$, we have
	\begin{equation}
		\frac{\log\mathcal{K}(t)}{N} = \int_{0}^{1}\int_{0}^{2\pi} \frac{u\,du\,d\theta}{\pi} \log\left[\sum_{k=0}^{L} \widetilde{g}_k \cos\left(kt\left(2u\cos\theta - \mu_R\right)\right)\right].
	\end{equation}
	To facilitate the analysis, we rewrite this expression as
	\begin{equation}\label{eq:coherent-leading-SFF}
		\begin{aligned}
			\frac{\log\mathcal{K}(t)}{N} &= \int_{0}^{1}\int_{0}^{2\pi} \frac{u\,du\,d\theta}{\pi} \log\left[1 + F(u\cos\theta)\right] \\
			&= \sum_{j=1}^{\infty} \int_{0}^{1}\int_{0}^{2\pi} \frac{u\,du\,d\theta}{\pi} \frac{(-1)^{j-1} \left(F(u\cos\theta)\right)^j}{j},
		\end{aligned}
	\end{equation}
	where we define
	\begin{equation}
		F(u\cos\theta) = -1 + \sum_{k=0}^L \widetilde{g}_k \cos\left(kt\left(2u\cos\theta - \mu_R\right)\right), \quad |F(u\cos\theta)| \leq 1.
	\end{equation}
	Using the integral identity
	\begin{equation}
		\int_{0}^{1}\int_{0}^{2\pi} \frac{u\,du\,d\theta}{\pi} \cos\left(kt\left(2u\cos\theta - \mu_R\right)\right) = \frac{J_1(2kt) \cos(k\mu_R t)}{kt},
	\end{equation}
	we can express $\frac{\log\mathcal{K}(t)}{N}$ as an expansion in terms of Bessel functions. For the case $z_R(x) = 1 + mx$, we obtain
	\begin{equation}
		\frac{\log\mathcal{K}(t)}{N} = \sum_{j=1}^{\infty} \sum_{k=0}^{j} \frac{(-1)^{j-1}}{j} \frac{r_1^j}{2^j}  \binom{j}{k} \cos\left[(j-2k)\mu_R t\right] \frac{2J_1(2|j-2k|t)}{2|j-2k|t},
	\end{equation}
	where $r_1 = \frac{2m}{m^2+1} \leq 1$. The formula in Eq.~\eqref{eq:coherent-leading-SFF} cannot produce the exponential ramp $\log \mathcal{K}(t) \sim C_0 t$ in the region $1 \ll t \ll N$, since it ultimately depends on terms of the form $\frac{J_1(2kt)\cos(k\mu_R t)}{kt}$. These terms have the asymptotic behavior
	\begin{equation}
		\left|\frac{J_1(2kt)}{kt}\right| \sim \frac{1}{t^{3/2}} \ll 1 \quad \text{for} \quad 1 \ll t \ll N, \quad k \neq 0.
	\end{equation}
	Consequently, the SFF obtained via the coherent state approach rapidly approaches the plateau in this time region. This conclusion remains valid even when considering higher-order approximations of $\text{tr}_N$, as the ramp structure may be lost in the approximation of Eq.~\eqref{eq:G_trace}.
	As a concrete example, consider the first-order correction (refer to Eq.~\eqref{eq:first-order})
	
	\begin{equation}
		\frac{(\log\mathcal{K}(t))^{(1)}}{N} = \int_{u_a}^{u_b} \int_{0}^{2\pi} \widetilde{\mathsf{w}}_1(u) \frac{u\,du\,d\theta}{\pi} \log\left[1 + F(u\cos\theta)\right],
	\end{equation}
	with integration bounds $u_a = r_a/\sqrt{N} = 1 - v_a/\sqrt{N}$ and $u_b = r_b/\sqrt{N} = 1 + v_b/\sqrt{N}$. The weight functions are
	\begin{equation}
		\begin{aligned}
			\widetilde{\mathsf{w}}_1(u_a < u < 1) &= \frac{\Gamma(N,N)}{\Gamma(N)} - \frac{2e^{-N}N^{N+\frac{1}{2}}(u\sqrt{N} - \sqrt{N})}{\Gamma(N+1)} - 1, \\
			\widetilde{\mathsf{w}}_1(1 < u < u_b) &= \frac{\Gamma(N,N)}{\Gamma(N)} - \frac{2e^{-N}N^{N+\frac{1}{2}}(u\sqrt{N} - \sqrt{N})}{\Gamma(N+1)}.
		\end{aligned}
	\end{equation}
	Using the integral identities
	\begin{equation}
		\begin{aligned}
			&\int u\,du \int_{0}^{2\pi} \frac{d\theta}{\pi} \cos\left(kt(2u\cos\theta - \mu_N)\right) = u^2 \cos(\mu_N kt) \,_{0}F_1(;2;-k^2t^2u^2), \\
			&\int u\,du \int_{0}^{2\pi} \frac{d\theta}{\pi} \cos\left(kt(2u\cos\theta - \mu_N)\right) = \frac{1}{3}u^3 \cos(\mu_N kt) \,_{1}F_2\left(\frac{3}{2};1,\frac{5}{2};-k^2t^2u^2\right),
		\end{aligned}
	\end{equation}
	we can express $\frac{(\log\mathcal{K}(t))^{(1)}}{N}$ in terms of hypergeometric functions $_0F_1(;2;-a^2t^2)$ and $_1F_2\left(\frac{3}{2};1,\frac{5}{2};-b^2t^2\right)$. Both these functions (and their derivatives) exhibit oscillatory behavior with decaying amplitude, rapidly approaching zero for $t\gg 1$. Similarly, one can show that higher-order corrections still do not contribute to the ramp behavior in the region $1 \ll t \ll N$.
	
	\subsection{Cluster function approach}
	The SFF can also be analyzed using the cluster function approach. We begin by defining
	\begin{equation}
		|z_{R}|^{2}(t) = g_{0}\left(1 + \sum_{k=1}^{L} r_{k}\cos\left(kt(\varepsilon-\mu)\right)\right) \equiv g_{0}\left(1 + F(\varepsilon-\mu,t)\right),
	\end{equation}
	where $r_k = g_k/g_0$ are the normalized coefficients. 
	An important identity emerges when considering products of $F$
	\begin{equation}
		\prod_{j=1}^{n}F(\varepsilon_{j},t) = 2^{-n}\sum_{\substack{\zeta_{j}=-L \\ \zeta_{j}\neq 0}}^{L} \left(\prod_{j=1}^{n} r_{|\zeta_{j}|}\right) e^{it\sum_{j=1}^{n}(\varepsilon_{j}-\mu)\zeta_{j}}.
	\end{equation}
	Following the methodology outlined in Appendix \ref{appendix:cluster}, we derive the SFF expression
	\begin{equation}
		\mathcal{K}(t) = \frac{g_0^N}{D^{2N}} \exp\left[N\sum_{k=1}^{L} r_{k}\frac{J_{1}(2kt)}{kt}\cos(k\mu t) + NA_{0}(t) + 2N\sum_{p=1}^{NL} A_{p}(t)\frac{\sin\left(\frac{\pi}{2}pt\right)}{\frac{\pi}{2}pt}\cos(p\mu t)\right],
	\end{equation}
	where the coefficients $A_p(t)$ are given by
	\begin{equation}
		A_{p}(t) = \sum_{n=2}^{N}(-1)^{n-1}\frac{1}{n2^{n}} \sum_{\sum\zeta_{i}=p}' r(\{\zeta_{i}\}) \left[1 - \frac{t}{2N}s(\{\zeta_{i}\})\right] \Theta\left[1 - \frac{t}{2N}s(\{\zeta_{i}\})\right]\ed
	\end{equation}
	Here $r(\{\zeta_{i}\})\equiv \prod_{j=1}^{n} r_{|\zeta_{j}|}$, the $s(\{\zeta_{i}\})$ is defined as
	\begin{equation}\label{eq:s-factor}
		s(\{\zeta_i\}) = \max\left\{0, \sum_{i=1}^j \zeta_i\right\}_{j=1}^{n-1} - \min\left\{0, \sum_{i=1}^j \zeta_i\right\}_{j=1}^{n-1}.
	\end{equation}
	We now analyze the SFF behavior in three distinct time regions:
	\begin{itemize}
		\item Short-to-Intermediate Times: $0 < t \ll N$
		
		For this region, we approximate $\Theta\left[1-\frac{t}{2N}s(\{\zeta_{i}\})\right] \approx 1$ in $A_p(t)$, yielding
		\begin{equation}
			A_{p}(t\ll N) \approx B_{p} + C_{p}\frac{t}{N},
		\end{equation}
		where the coefficients are
		\begin{equation}\label{eq:BpCp}
			\begin{aligned}
				B_{p} = \sum_{n=2}^{N}\frac{(-1)^{n-1}}{n2^{n}}\sum'_{\sum\zeta_{i}=p}r(\{\zeta_{i}\}), ~
				C_{p} = \sum_{n=2}^{N}\frac{(-1)^{n}}{n2^{n+1}}\sum'_{\sum\zeta_{i}=p}r(\{\zeta_{i}\})s(\{\zeta_{i}\}).
			\end{aligned}
		\end{equation}
		
		The SFF expression becomes
		\begin{equation}
			\mathcal{K}_{t\ll N} = \frac{g_{0}^{N}}{D^{2N}}\exp\left[
			\begin{aligned}
				&N\sum_{k=1}^{L}r_{k}\frac{J_{1}(2kt)}{kt}\cos(k\mu t) + N(B_{0}+C_{0}t/N) \\
				&+ 2N\sum_{p=1}^{NL}(B_{p}+C_{p}t/N)\frac{\sin(\frac{\pi}{2}pt)}{\frac{\pi}{2}pt}\cos(p\mu t)
			\end{aligned}
			\right].
		\end{equation}
		For physical consistency, we require
		\begin{equation}
			\lim_{N\to\infty}B_p = \text{finite}, \quad \lim_{N\to\infty}C_p/N = \text{finite}.
		\end{equation}
		\begin{itemize}
			\item Early Time: $0 < t \ll 1$
			
			Neglecting $C_p$ terms ($C_pt/N \ll B_p$), we obtain
			\begin{equation}
				\mathcal{K}_{t\ll1} = \frac{g_{0}^{N}}{D^{2N}}\exp\left[
				NB_{0} + N\sum_{k=1}^{L}r_{k}\frac{J_{1}(2kt)}{kt}\cos(k\mu t) + 2N\sum_{p=1}^{\infty}B_{p}\frac{\sin(\frac{\pi}{2}pt)}{\frac{\pi}{2}pt}\cos(p\mu t)
				\right].
			\end{equation}
			The normalization condition $\mathcal{K}(0)=1$ imposes
			\begin{equation}\label{eq:Bp_sum}
				2\sum_{p=1}^{\infty}B_{p} = \log\left(1+\sum_{k=1}^{L}r_{k}\right) - \sum_{k=1}^{L}r_{k} - B_{0}.
			\end{equation}
			To second order in $t$
			\begin{equation}\label{eq:SFF-earlytime-cluster}
				{\log \mathcal{K}_{t\ll1}\over N} = - \left[\left(\mu^{2}+\frac{\pi^{2}}{12}\right)\sum_{p=1}^{\infty}p^{2}B_{p} + \frac{\mu^{2}+1}{2}\sum_{k=1}^{L}k^{2}r_{k}\right]t^{2} + \mathcal{O}(t^{4}).
			\end{equation}
			\item Intermediate Time: $1 \ll t \ll N$
			
			Dominant contribution comes from $A_0(t)$
			\begin{equation}\label{eq:middle-time}
				\mathcal{K}_{1\ll t\ll N} = \frac{g_{0}^{N}}{D^{2N}}\exp\left[
				N\sum_{k=1}^{L}r_{k}\frac{J_{1}(2kt)}{kt}\cos(k\mu t) + NB_{0} + C_0t
				\right].
			\end{equation}
			For $t \gg t_p \equiv (N/C_0)^{2/5}$, the exponential ramp dominates
			\begin{equation}
				\mathcal{K}_{t_p\ll t\ll N} = \frac{g_{0}^{N}}{D^{2N}}\exp\left[NB_{0} + C_0t\right].
			\end{equation}
		\end{itemize}

		\item Late Time: $t \gg N$
		
		The step function vanishes, leaving
		\begin{equation}
			\mathcal{K}_{t\gg N} = \frac{g_0^N}{D^{2N}}\exp\left[N\sum_{k=1}^{L}r_{k}\frac{J_{1}(2kt)}{kt}\cos(k\mu t)\right].
		\end{equation}
		
		At infinite time, we have $\mathcal{K}_{t=\infty} = \frac{g_0^N}{D^{2N}}$, it is 
		consistent with the result of  coherent state approach (Eq.~\eqref{eq:coherence-plateau}).
		
	\end{itemize}
	
	\subsubsection{$R$-para-fermions with $z_R(x)=1+mx$}
	\label{subsec:analytical-exp}
	When \( m = 1 \) (or $r_1=1$ in this subsection), the system is reduced to the case of ordinary fermions, which has been extensively discussed in the literature \cite{Liao_2020}. In this section, we consider the generalization to arbitrary \( m \). To facilitate comparison with previous studies, we adopt the same notation, such as the conventions for the \( B \) and \( C \) coefficients. It is direct to find 
	\begin{align}
		|z_{R}|^2(t)=m^{2}+1+2m\cos\left( t(\varepsilon-\mu)\right),
	\end{align}
	then $g_{0}=m^{2}+1,g_{1}=2m,D=m+1$, so that we have 
	\begin{align}
		r_{1}=\frac{2m}{m^{2}+1}\le 1,~\xi_i=\pm 1\ed
	\end{align}
	Using the definition in Eq.~\eref{eq:BpCp}, we have
	\begin{align}\label{eq:Bp-exact}
		B_{0}&=\sum_{n=\text{even},n\ge2}\frac{\left(-1\right)^{n-1}r_1^n}{n2^{n}}\binom{n}{{n\over 2}}=\log\left(\frac{1}{2}\left(\sqrt{1-r_{1}^{2}}+1\right)\right)\co\nn
		B_{1}&=\sum_{n=\text{odd},n\ge2}\frac{\left(-1\right)^{n-1}r_1^n}{n2^{n}}\binom{n}{{n-1\over 2}}=\frac{2-r_{1}^{2}-2\sqrt{1-r_{1}^{2}}}{2r_{1}}\ed
	\end{align}
	For $k>1$, we have a universal relation 
	\begin{align}
		B_{k}=\frac{(-1)^{k-1}}{k}\left[\frac{\cos\left(\frac{\theta_{1}}{2}\right)-\sin\left(\frac{\theta_{1}}{2}\right)}{\cos\left(\frac{\theta_{1}}{2}\right)+\sin\left(\frac{\theta_{1}}{2}\right)}\right]^{k},~r_1=\cos \theta_1\ed
	\end{align}
	Then one can check the results are consistent with Eq. \eref{eq:Bp_sum}. Moreover, they return back to the results in Appendix of \cite{Liao_2020} when $r_1=1$ for ordinary fermions.  For $t\ll 1$, we have 
	\begin{align}
		{\log \mathcal{K}_{t\ll 1}\over N}=-\left[\frac{\left(\mu^{2}+1\right)m}{(m+1)^{2}}+\frac{\left(12-\pi^{2}\right)m^{2}}{6(m+1)^{2}\left(m^{2}+1\right)}\right]t^2+\mathcal{O}(t^4)\co
	\end{align}
	while the coherent state approach gives 
	\begin{align}
		{\log \mathcal{K}_{t\ll 1}\over N}=-\frac{\left(\mu^{2}+1\right)m}{(m+1)^{2}}t^2+\mathcal{O}(t^4)\ed
	\end{align}
	The numerical simulation indicate the coherent state approach is exact for short times, while the cluster approach with box approximation is not accurate. 
	For $C_0$, following the derivations in Appendix of \cite{Liao_2020}, we have 
	\begin{align}
		C_{0}=\sum_{n=\text{even},n\ge 2}^{N}\frac{(-1)^{n}r_1^n}{n2^{n+1}}\left(2^{n}-\frac{n!}{\left(\frac{n}{2}\right)!\left(\frac{n}{2}\right)!}\right)\ed
	\end{align}
	So finally we obtain
	\begin{equation}\label{eq:C0-exact}
		\begin{aligned}
			C_{0}=\begin{cases}
				-\frac{1}{4}\log\left(1-r_{1}^{2}\right)+\frac{1}{2}\log\left(\frac{1}{2}\left(\sqrt{1-r_{1}^{2}}+1\right)\right), & r_{1}<1\co\\
				\frac{1}{4}\left(\ln\frac{N}{8}+\gamma_{E}\right)\approx 0.25 \ln N-0.375556, & r_{1}=1\co
			\end{cases}
		\end{aligned}
	\end{equation}
	where $\gamma_E \approx  0.577$ is the Euler-Mascheroni constant. 
	In terms of $m$, we obtain a simplified expression
	\begin{align}\label{eq:exB-C0-m}
		C_{0,m>1} = \frac{1}{2}\log\left(\frac{m^{2}}{m^{2}-1}\right).
	\end{align}
	Numerically, for the case $L=1$, we derive an alternative formulation using generating functions. By observing that
	\begin{align}
		\sum_{\substack{\sum \zeta_i = 0}}' r(\{\zeta_i\}) = Q(x)^n \Big\vert_{x^0}, ~
		\sum_{\substack{\sum \zeta_i = 0}} r(\{\zeta_i\}) s(\{\zeta_i\}) = 2\sum_{l=1}^{\lfloor n/2 \rfloor} Q(x)^n \Big\vert_{x^{2l}},
	\end{align}
	where $Q(x) \equiv r_1\left(x + \frac{1}{x}\right)$, we establish a connection between the combinatorial sums and the generating function $Q(x)$.
	\begin{figure}[ht]
		\begin{center}
			\includegraphics[width=0.45\textwidth]{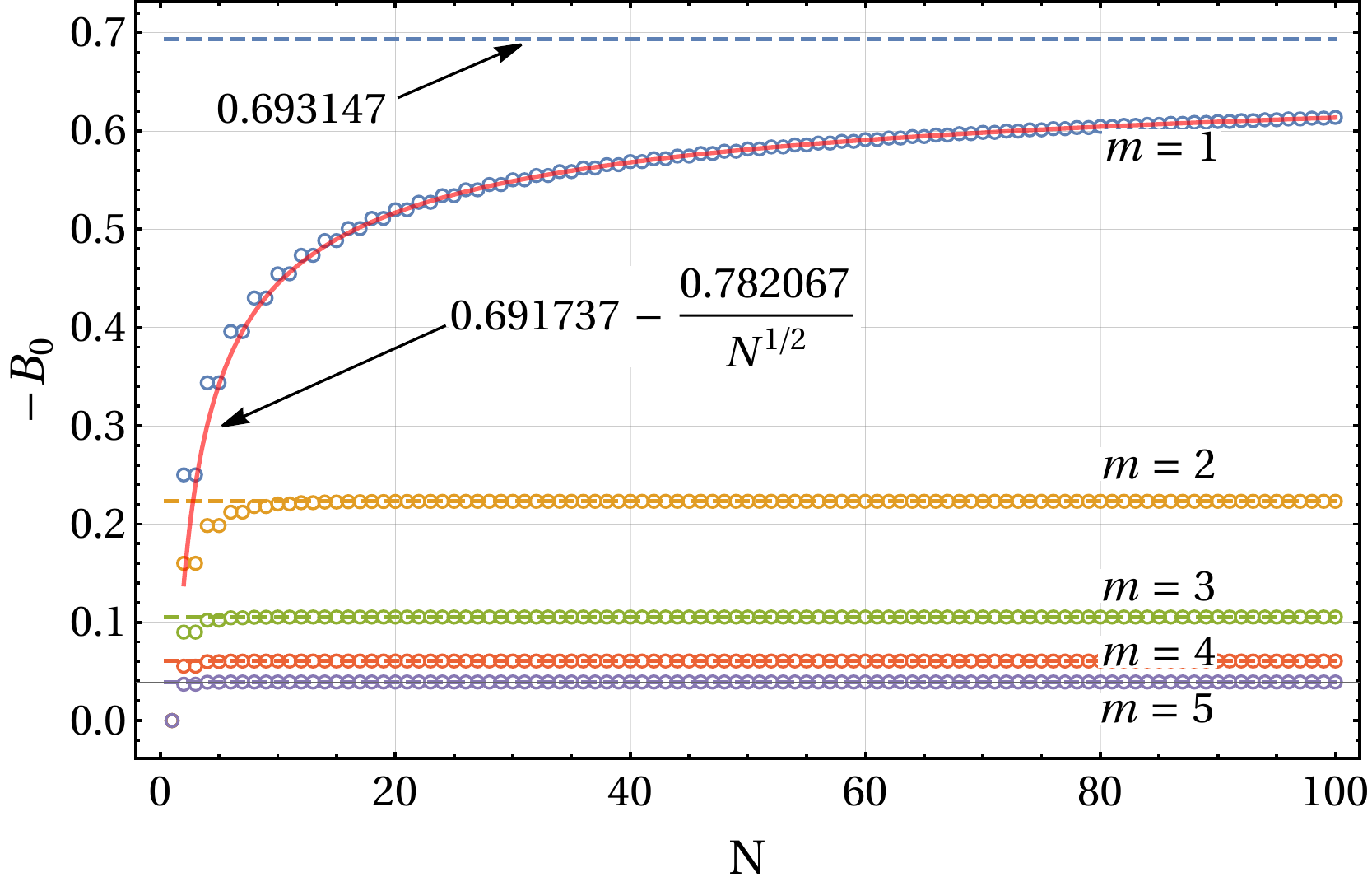}
			\includegraphics[width=0.45\textwidth]{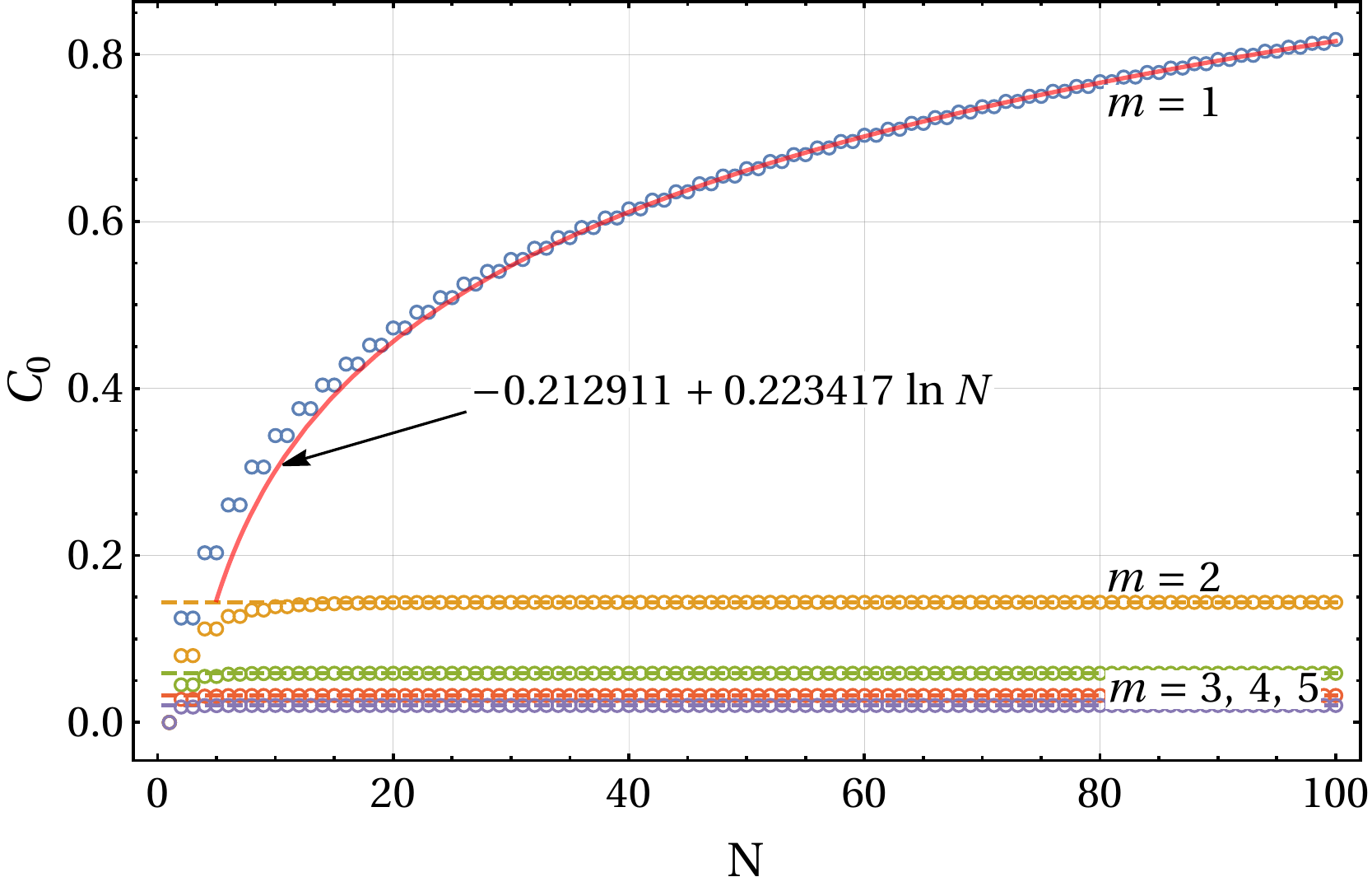}
			\caption{The plot shows $-B_0$ and $C_0$ computed numerically as functions of $N$ for $z_R(x) = 1 + m x$. The dashed lines represent the exact analytical results derived from Eq.~\eqref{eq:Bp-exact} and Eq.~\eqref{eq:exB-C0-m}. Focusing on the asymptotic regime, we restrict the fitting to data with $N \geq 20$. The numerical results exhibit excellent agreement with the exact theoretical predictions.}
			\label{fig:B0C0-exA}
		\end{center}
	\end{figure}
	For systems with $L > 2$, the transfer matrix method developed in Appendix~\ref{appdix:transferT} becomes essential for our analysis. Figure~\ref{fig:B0C0-exA} demonstrates excellent agreement between numerical computations and exact theoretical predictions, validating our analytical approach.
	
\begin{figure}[ht]
		\begin{center}
			\includegraphics[width=0.45\textwidth]{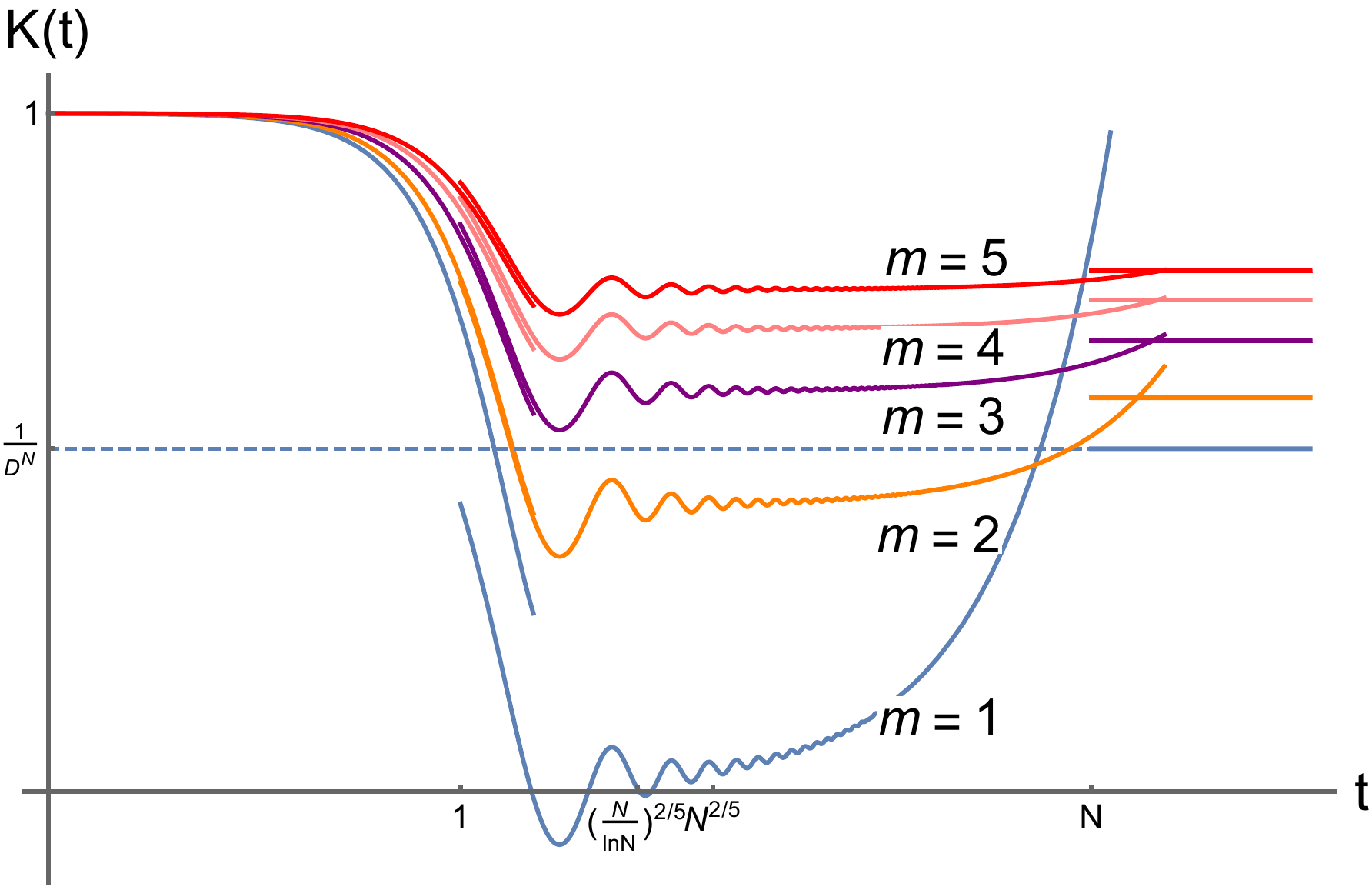}
			\includegraphics[width=0.4\textwidth]{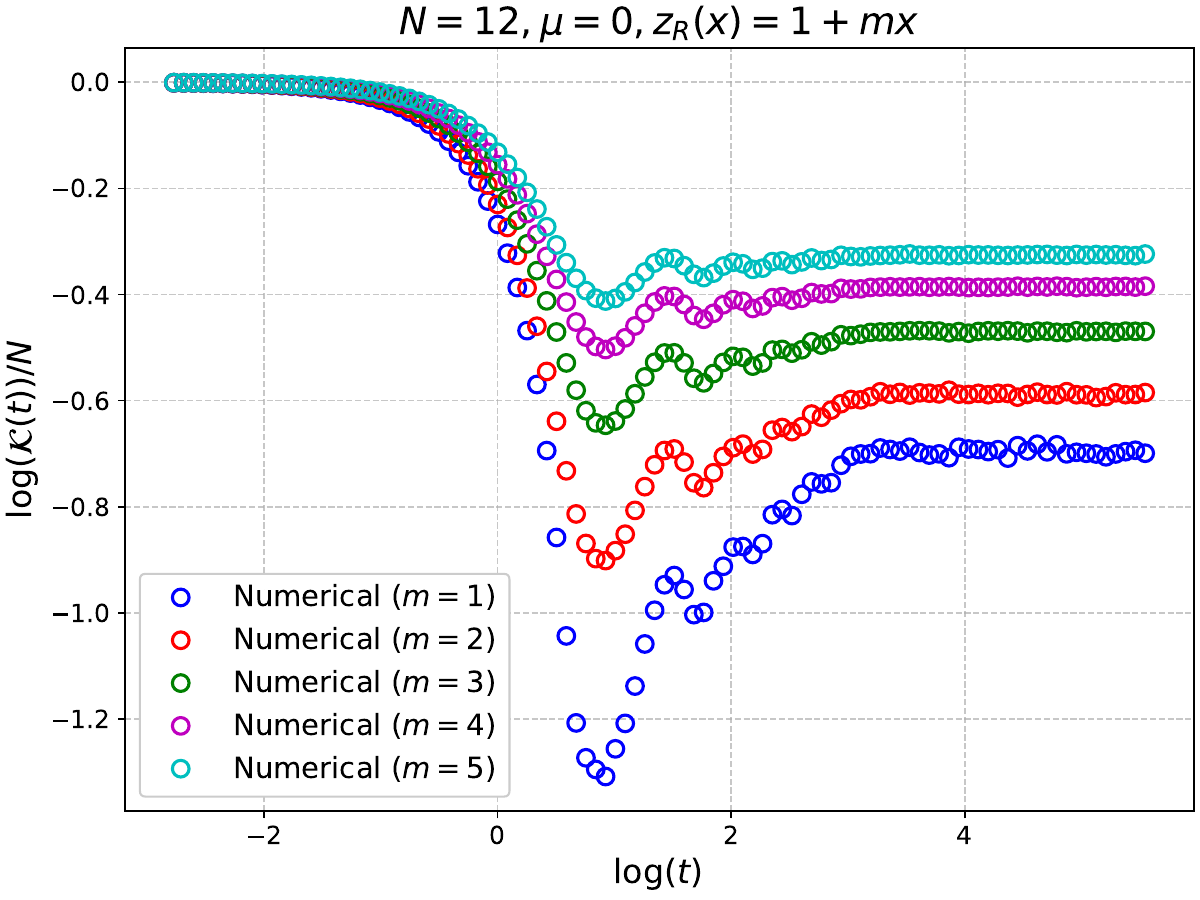}
			\caption{\textbf{Left}: Log-log plot of SFF using the cluster function approach for $z_R(x)=1+mx$ with $m=1,2,3,4,5$ ($N=400$). Time range $t\in[0.02,8N]$. The $m=1$ case (ordinary fermions) shows an exponential ramp with $C_0 \sim \mathcal{O}(\ln N)$, while $m>1$ cases exhibit qualitatively different behavior with $C_0 \sim \mathcal{O}(1)$. \textbf{Right}: Numerical simulation for $N=12$ with 20,000 samples ($t\in[2^{-4},2^8]$). While the small system size prevents observation of the exponential ramp, the plateau location agrees with the theoretical prediction $\mathcal{K}(t\to\infty) = (m^2+1)^{N}/(m+1)^{2N}$. }
			\label{fig:SFF}
		\end{center}
	\end{figure}

	A dramatic transition in the SFF emerges when comparing the cases $m=1$ and $m>1$, as illustrated in Fig.~\ref{fig:SFF} through both theoretical and numerical results. 
	The numerical analysis faces inherent limitations due to the system's scaling properties. The Hilbert space dimension grows as $D^N$, while the energy scales as $O(N)$. Consequently, the typical level spacing $\sim O(N/D^N)$ becomes exponentially small. When this spacing falls below machine precision ($\delta \approx 10^{-16}$), the GUE simulation breaks down. This constraint leads to the scaling relation
	\begin{align}
		\frac{N}{D^{N}} = \delta \Rightarrow N \approx \log_{D}(1/\delta),
	\end{align}
	which for $D=6$ yields $N \approx 20$ as the practical upper limit. Therefore, direct numerical comparison with the large-$N$ theoretical predictions presented in the main text becomes infeasible. Recently, a SYK$_2$-like model with couplings drawn from circular unitary ensembles (CUE) was studied in~\cite{Michael_2024}, where exact results for SFF were derived. Since CUE and GUE share the same kernel functions in the large-$N$ limit, we expect their results can be used as a benchmark. This connection will be explored in our forthcoming work.
	\subsubsection{$R$-para-fermions with $z_R(x)=1+mx+x^2$}
	Then we turn to deal with Example B, it is straightforward to obtain
	\begin{align}
		|z_R|^2(t)=m^{2}+2+4m\cos( t(\varepsilon-\mu))+2\cos(2t(\varepsilon -\mu))\co
	\end{align}
	so that 
	\begin{align}
		r_{1}=\frac{4m}{m^{2}+2},r_{2}=\frac{2}{m^{2}+2},\zeta_i=\pm 1,\pm 2\ed
	\end{align}
	In contrast to the previous case, here we may encounter situations where $r_1 > 1$. While determining the exact analytical expressions for $B_0$ and $C_0$ proves challenging, we can evaluate their scaling behavior numerically. These coefficients are defined as:
	\begin{equation}
		\begin{aligned}
			B_{0} = \sum_{n=2}^{N}\frac{(-1)^{n-1}}{n2^{n}}\sum'_{\substack{\sum\zeta_{i}=0}} r(\{\zeta_{i}\}),~
			C_{0} = \sum_{n=2}^{N}\frac{(-1)^{n}}{n2^{n+1}}\sum'_{\substack{\sum\zeta_{i}=0}} r(\{\zeta_{i}\})s(\{\zeta_{i}\}).
		\end{aligned}
	\end{equation}
	We can interpret these expressions through a path integral formulation. Defining $x_j = \sum_{i=1}^{j} \zeta_i$ with $x_0 = 0$, each set $\{\zeta_i\}$ corresponds to a path $\{x_i\}$. Thus, $B_0$ and $C_0$ represent sums over weighted paths constrained to return to the origin ($x_n = 0$). At each step $j$, the possible displacements $\zeta_{j+1} \in \{\pm1, \pm2, \ldots, \pm L\}$ have weights $r_{|\zeta_{j+1}|}$, giving each path a total weight of $\prod_{j=1}^n r_{\zeta_j}$. For $C_0$, we include an additional factor $s(\{\zeta_i\})$, which from Eq.~\eqref{eq:s-factor} corresponds to the path's vertical span: $\max(\{x_i\}) - \min(\{x_i\})$.
	The computation of $B_0$ can be approached either through the transfer matrix method or via generating functions. The latter approach utilizes the identity
	\begin{align}
		\sum_{\substack{\sum\zeta_{i}=p}}' r(\{\zeta_{i}\}) &= Q(x)^n \Big\vert_{x^p}, \\
		\text{where } Q(x) &= r_{1}(x + x^{-1}) + r_{2}(x^{2} + x^{-2})
	\end{align}
	where $\vert_{x^{p}}$ denotes extracting the coefficient of $x^p$ in the series expansion.
	Unlike the former case, here one may have  $r_1>1$. We just need to evaluate the scale of $B_0,C_0$,  It is hard to evaluate the $B_0,C_0$ analytically, so we may better consider the numerical method. Recall their definitions
	\begin{equation}
		\begin{aligned}
			B_{0} = \sum_{n=2}^{N}\frac{(-1)^{n-1}}{n2^{n}}\sum'_{\sum\zeta_{i}=0}r(\{\zeta_{i}\}), ~
			C_{0} = \sum_{n=2}^{N}\frac{(-1)^{n}}{n2^{n+1}}\sum'_{\sum\zeta_{i}=0}r(\{\zeta_{i}\})s(\{\zeta_{i}\}).
		\end{aligned}
	\end{equation}
	\begin{figure}[h]
		\begin{center}
			\includegraphics[width=0.45\textwidth]{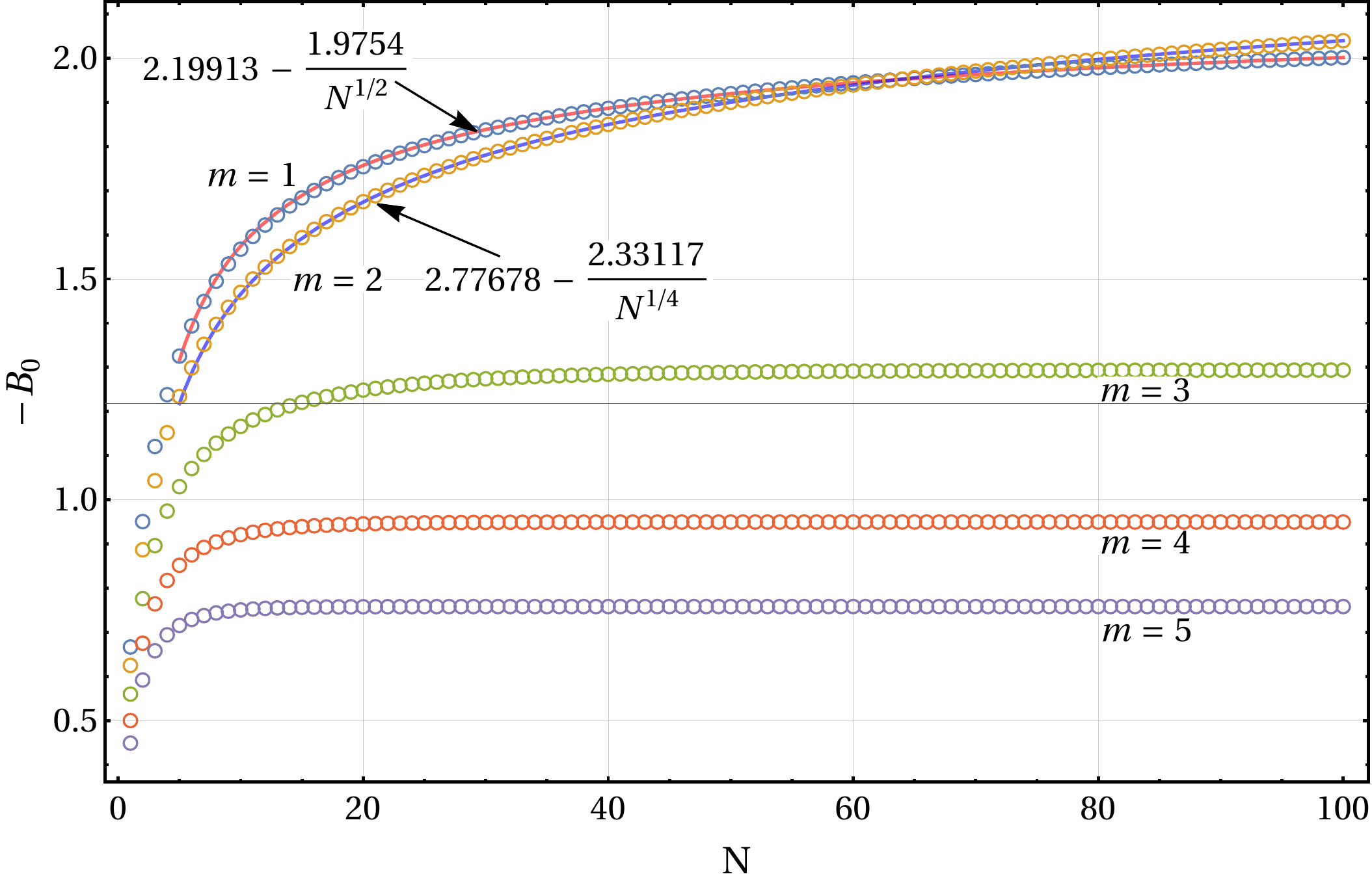}
			\includegraphics[width=0.45\textwidth]{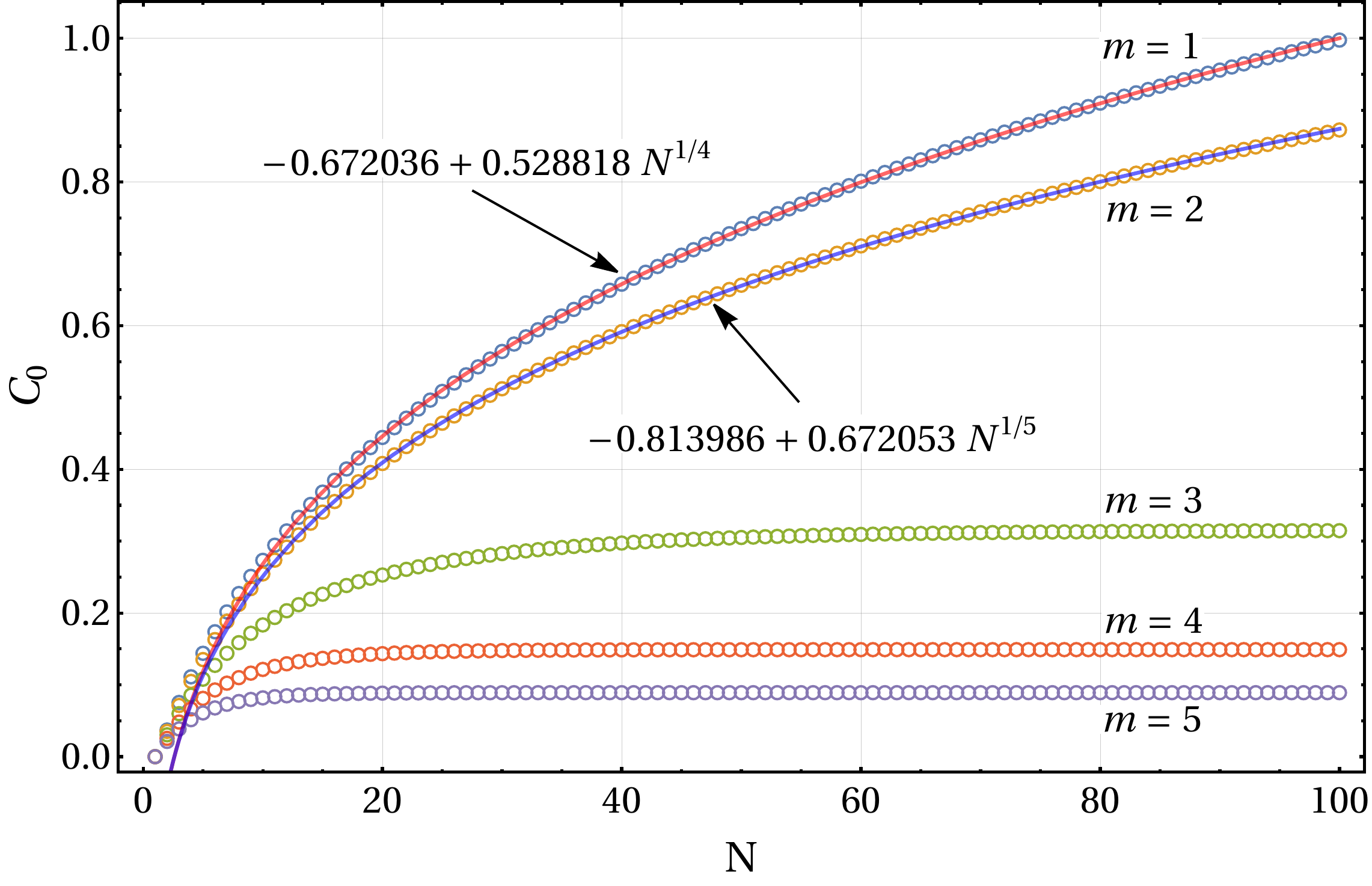}
			\caption{The plot of $-B_0,C_0$ with respect to $N$ for $z_R(x)=1+mx+x^2$.  Since we are concerned with asymptotic behavior, we use the data $N\ge 10$ to fit. We find that for $m=1,2$, $C_0$ is divergent with a power law in the large $N$ limit. Unlike the log divergence of $z_R(x)=1+mx$. As for $m\ge 2$, $C_0$ is finite.   }
			\label{fig:B0C0}
		\end{center}
	\end{figure}
	For our analysis, we employ the transfer matrix method developed in Appendix~\ref{appdix:transferT}, with the weight assignments $\mathsf{g}_0 = 0$ and $\mathsf{g}_k = r_k$. However, when $r_1 > 1$ (occurring for small $m$) and $r_2 > 0$, we encounter numerical convergence issues in the direct evaluation of $B_0$ and $C_0$ sums.
	To address this, we introduce an alternative expansion scheme. Defining $|z_{R}|^{2}(t) = D^2(1 + F(\varepsilon,t))$, we proceed with the following formulation
	\begin{equation}
		\begin{aligned}
			\prod_{j=1}^{n}F(\varepsilon_{j},t)&=\prod_{j=1}^{n}\left[\widetilde{g}_{0}-1+\sum_{k=1}^{L}\widetilde{g}_{k}\cos\left(kt\left(\varepsilon_{j}-\mu\right)\right)\right]\\&=\frac{1}{2^{n}}\sum_{\{\zeta\}=-L}^{L}\mathsf{g}_{\{\zeta\}}\exp\left(it\sum_{j=1}^{n}\zeta_{j}\left(\varepsilon_{j}-\mu\right)\right)
		\end{aligned}
	\end{equation}
	where we have defined 
	\begin{align}
		\mathsf{g}_{0}=2\left(\widetilde{g}_{0}-1\right),\mathsf{g}_{k>0}=\widetilde{g}_{k}\ed
	\end{align}
	Following an analogous derivation, we obtain the SFF
	\begin{align}
		\mathcal{K}(t) = \exp\left[N\mathsf{g}_0/2 + N\sum_{k=1}^{L}\mathsf{g}_{k}\frac{J_{1}(2kt)}{kt}\cos\left(k\mu t\right) + NA_{0}(t) + 2N\sum_{p=1}^{NL}A_{p}(t)\frac{\sin\left(\frac{\pi}{2}pt\right)}{\frac{\pi}{2}pt}\cos\left(p\mu t\right)\right],
	\end{align}
	where $A_p(t) = B_p + \frac{1}{N}C_p t$ retains its previous form but now includes $\zeta_i = 0$ as a possible value in the sums. 
	The numerical implementation benefits from two key conditions: (i) $\mathsf{g}_0 < 0$ with $|\mathsf{g}_0| < 2$, and (ii) $0 < \mathsf{g}_k < 1$ for $k \neq 0$, which together ensure rapid convergence. The modified coefficients are defined as
	\begin{equation}
		\begin{aligned}\label{eq:BpCp-for-num}
			B_{p} = \sum_{n=2}^{N}\frac{(-1)^{n-1}}{n2^n}\sum_{\substack{\sum\zeta_{i}=p}}\mathsf{g}(\{\zeta_{i}\}), ~
			C_{p} = \sum_{n=2}^{N}\frac{(-1)^{n}}{n2^{n+1}}\sum_{\substack{\sum\zeta_{i}=p}}\mathsf{g}(\{\zeta_{i}\})s(\{\zeta_{i}\}),
		\end{aligned}
	\end{equation}
	where $\zeta_i$ now takes values in $\{-L,\ldots,0,\ldots,L\}$, and consequently the prime has been removed from the summation symbol. 
	For numerical computation, we evaluate $B_0$ and $C_0$ using the transfer matrix approach detailed in Appendix~\ref{appdix:transferT}. Although Eq.~\eqref{eq:BpCp-for-num} differs in form from Eq.~\eqref{eq:BpCp}, both formulations must yield equivalent physical predictions. This consistency requires, in particular, that the values of $C_0$ obtained from both definitions agree with each other.
	
	The remaining task is to determine the asymptotic behavior of $B_0$ and $C_0$ for large $N$ using numerical data. It is expected that $B_0$ is always convergent for large $N$, while for some cases, $C_0$ can be divergent as $N$ goes to infinity. We assume the divergent behavior is $\log N$. Denote $c_N$ as a numerical list of $N$ (taking $c_N$ to be $B_0$ or $C_0$). We test convergence by fitting \( f(N) = a + b\log N \) to \( N > N_* \gg 1 \). 
	Large residuals for \( N < N_* \) indicate model mismatch, prompting power-law fits \( f(N) = a + b N^\gamma \) (\(\gamma > 0\)). For the convergent $c_N$ that approaches its limit exponentially ($c_N\sim c_{\infty}+ae^{-bN},b>0$), we numerically take the approximation $c_{\infty}\approx c_{N_0}$ with a large but finite $N_0\gg 1$. If $c_N$ approaches its limit polynomially ($c_N\sim c_{\infty}+aN^{-b},b>0$), we first fit $c_N$ with data for $N\ge N_*$, then take the limit $N\to \infty$ to obtain $c_{\infty}$, where a proper $b$ is chosen so that the fitting function has a small error for the data near $N< N_*$. 
	
	The numerical results for Example B are presented in Fig.~\ref{fig:B0C0}. While we omit the calculation of higher-order coefficients $B_p$ and $C_p$ (as they do not affect the ramp behavior), our analysis reveals two key features: (i) $B_0$ remains convergent in all cases, while (ii) $C_0$ exhibits power-law divergence for $m=1,2$. 
	
	This behavior is particularly notable when compared with the trivial $R$-PSYK$_2$ case ($L=2$) described by $z_R(x)=(1+x)^2=1+2x+x^2$. These observations motivate the following conjecture: For a general $R$-PSYK$_2$ model with generating function $z_R(x)=\sum_{n=0}^L d_n x^n$ ($d_0=d_L=1$), the exponential ramp displays a divergent growth rate $C_0$ in the large-$N$ limit if and only if $z_R(x)\le (1+x)^L$ for all $x>0$. Otherwise, $C_0$ remains $\mathcal{O}(1)$. 
	Currently, we have numerically verified this conjecture for Examples A and B. A rigorous analytical proof, however, would require exact evaluation of the $C_0$ expression, which remains an open mathematical challenge.

	\section{Time Evolution }
	\label{sec:time-evolution}
	In this section, we examine the time evolution of $\psi_{i,a}^{\pm}$, which is essential for computing correlation functions. The exchange statistics between operators at different sites proves nontrivial, as each $\psi_{i,a}$ constitutes a global operator representable as a sum of MPO string operators acting on a spin chain. These operators $\psi_{i,a}^{\pm}$ act non-trivially on the first $i$ spins.
	
	Nevertheless, we can directly evaluate the time evolution. A more effective approach involves constructing the basis through local spin operators $x_{i,a}^{\pm}$ and $y_{i,a}^{\pm}$, where $x_{i,a}^{\pm}$ satisfies distinct commutation relations (see Eq.~(S21) in the arXiv version of \cite{Wang_2025}). For general $R$, we have
	\begin{eqnarray}\label{eq:JWT_string}
		\hat{\psi}_{ia}^-&=&\begin{tikzpicture}[baseline={([yshift=.4ex]current bounding box.center)}, scale=0.64]
			\node  at (-1.5*\AL,0.*\AL) {\footnotesize $a$};
			\Tmmatrix{0}{0}{}
			\node  at (0,-1.6*\AL) {\footnotesize $1$};
			\Tmmatrix{2*\AL}{0}{}
			\node  at (2*\AL,-1.6*\AL) {\footnotesize $2$};
			\Tmmatrix{4*\AL}{0}{}
			\node  at (4*\AL,-1.6*\AL) {\footnotesize $3$};
			\draw[dotted, thick] (5*\AL, 0) -- (7*\AL,0);
			\Tmmatrix{8*\AL}{0}{}
			\node  at (8*\AL,-1.6*\AL) {\footnotesize $i-1$};
			\ytriangle{10*\AL}{0}{-}{}
			\node  at (10*\AL,-1.6*\AL) {\footnotesize $i$};
		\end{tikzpicture},\nonumber\\
		\hat{\psi}_{ia}^+&=&\begin{tikzpicture}[baseline={([yshift=.4ex]current bounding box.center)}, scale=0.64]
			\node  at (-1.5*\AL,0.*\AL) {\footnotesize $a$};
			\Tpmatrix{0}{0}{}
			\node  at (0,-1.6*\AL) {\footnotesize $1$};
			\Tpmatrix{2*\AL}{0}{}
			\node  at (2*\AL,-1.6*\AL) {\footnotesize $2$};
			\Tpmatrix{4*\AL}{0}{}
			\node  at (4*\AL,-1.6*\AL) {\footnotesize $3$};
			\draw[dotted, thick] (5*\AL, 0) -- (7*\AL,0);
			\Tpmatrix{8*\AL}{0}{}
			\node  at (8*\AL,-1.6*\AL) {\footnotesize $i-1$};
			\ytriangle{10*\AL}{0}{+}{}
			\node  at (10*\AL,-1.6*\AL) {\footnotesize $i$};
		\end{tikzpicture},
	\end{eqnarray}
	where $\hat{y}^\pm_{ja}\equiv\!\!\begin{tikzpicture}[baseline={([yshift=-.2ex]current bounding box.center)}, scale=0.64]
		\ytriangle{0}{0}{\pm}{}
		\node  at (0,-1.6*\AL) {\footnotesize $ j$};
		\node  at (-1.4*\AL,0) {\footnotesize $a$};
	\end{tikzpicture}\!\!$, and $\hat{T}^\pm_{j,ab}\equiv\!\!\begin{tikzpicture}[baseline={([yshift=-.2ex]current bounding box.center)}, scale=0.64]
		\Tpmmatrix{0}{0}{}
		\node  at (0,-1.6*\AL) {\footnotesize $j$};
		\node  at (-1.4*\AL,0) {\footnotesize $a$};
		\node  at (1.4*\AL,0) {\footnotesize $b$};
	\end{tikzpicture}\!\!=\mp [\hat{y}^\pm_{j,a},\hat{x}^\mp_{j,b}]$ 
	are local spin operators acting on site $j$. Both $\hat{\psi}_{i,a}^\pm$ act non-trivially on sites $1,2,\ldots,i$ and act as identity on the rest of the chain.  
	Explicitly, we have 
	\begin{align}\label{eq:psi-string}
		\psi_{i,a}^{\pm}=T_{1,ab_{1}}^{\pm}\otimes T_{2,b_{1}b_{2}}^{\pm}\otimes \ldots T_{i-1,b_{i-2}b_{i-1}}^{\pm}\otimes y_{i,b_{i-1}}^{\pm}=\left(\otimes_{k=1}^{i-1}T_k^{\pm}\right)_{ab}\otimes y_{i,b}^{\pm}
	\end{align}   
	where the summation over $b_i$ is omitted. For a local operator $\hat{O}_k$ acting on the local Hilbert space of site $k$, we  its time evolution is given by
	\begin{equation}
		\hat{O}_{k}(\beta+it)=e^{\left(-\beta+it\right)H}\hat{O}_{k}e^{\left(-\beta-it\right)H},
	\end{equation}
	with 
	\begin{align}
		H=\sum_{a=1}^{m}\sum_{k=1}^{N}\left(\frac{\hat{x}_{k,a}^{+}\hat{y}_{k,a}^{-}+\hat{x}_{k,a}^{-}\hat{y}_{k,a}^{+}}{2}\varepsilon_{k}-\mu\hat{y}_{k,a}^{+}\hat{y}_{k,a}^{-}\right)\equiv \sum_{k=1}^N H_k\ed 
	\end{align}
	Since $[H_i,H_j]=0,\forall i\not=j$, it is easy to obtain the time evolution of a local operator via
	\begin{align}
		\hat{O}_{k}(\beta+it)=e^{\left(-\beta+it\right)H_k}\hat{O}_ke^{\left(-\beta-it\right)H_k}\ed
	\end{align}
	Note that all operators on the right-hand side of Eq.~\eref{eq:psi-string} are local, making their time evolution straightforward to compute, since their matrix elements are known. With the time evolution of the building blocks $\psi_{i,a}^{\pm}$ established, we can compute various correlation functions, including two-point functions and OTOCs (see \cite{syk2open} for the fermionic SYK$_2$ case). However, due to their complicated structure, performing the ensemble average may prove difficult. We omit explicit calculations here.
	\section{Conclusion}
	\label{sec:conclusion}
	
	In this paper, we conduct a comprehensive investigation of the $q=2$ para-Sachdev-Ye-Kitaev model ($R$-PSYK$_2$), analyzing both its thermodynamic properties and spectral form factor (SFF). Our results demonstrate that the coherent state approach provides an effective framework for studying finite-temperature behavior, while the cluster function approach with box approximation fails even at high temperatures. Moreover, through the coherent state formalism, we prove the self-averaging property of general $R$-PSYK$_2$ models in the large $N$ limit. While our focus has not been on correlation functions, their calculation follows directly from the methods developed in Section~\ref{sec:time-evolution}. Future studies could investigate how $R$-para-particle statistics influence two-point functions and out-of-time-ordered correlators (OTOCs).

	Our analysis of the SFF reveals a notable transition in its ramp behavior. For the case $z_R(x) = 1 + mx$, we find analytically that $C_0 = \mathcal{O}(1)$ when $m > 1$, contrasting with the $C_0 = \mathcal{O}(\ln N)$ scaling observed in conventional SYK$_2$ models. Numerical studies of the case $z_R(x) = 1 + mx + x^2$ indicate that $C_0 = \mathcal{O}(1)$ for $m\ge 3$ while $C_0\sim \mathcal{O}(N^\alpha),0<\alpha<1$ for  $m = 1, 2$. These results suggest that such transitions in $C_0$ may represent a universal characteristic of $R$-PSYK$_2$ systems, though other possible scaling behaviors remain to be explored. Mathematically, $C_0$ can be interpreted as a sum of weighted paths in Eq.~\eqref{eq:BpCp}. For $z_R(x) = 1 + mx$, the negative weight $r_1 < 0$ suggests convergence of $C_0$ as $N \to \infty$, though rigorous proof remains outstanding. Analytical treatment becomes increasingly difficult for higher-order $z_R(x)$, leading us to employ numerical methods based on Eq.~\eqref{eq:BpCp-for-num}. The slow growth of $C_0$ with $N$ makes precise determination of scaling laws challenging, leaving open possibilities such as $C_0 \sim N^\alpha$ ($0<\alpha < 1$) or $C_0 \sim N^\alpha \ln N$.
	
	The observed dramatic change in the scaling of $C_0$ at large $N$ currently lacks a complete physical interpretation. Building on the gravitational connection established in \cite{Saad:2018bqo}, where the SYK SFF relates to double cone solutions, we note that our SFF computation involves only local quantities. This local nature allows for the replacement of $\psi_{i,a}^{\pm}$ with local spin operators $\hat{y}_{i,a}^{\pm}$ in the Hamiltonian. Note that the commutation relations satisfied by the spin operators here involve $R$, so we cannot directly apply the fermionic path integral scheme. Perhaps we could generalize the Grassmann numbers used in fermionic path integrals to include flavor degrees of freedom and consistent with the commutation relations. Another brute-force approach might be to express the spin operators in terms of local auxiliary fermionic creation and annihilation operators, but this could introduce additional redundancies, thereby complicating the analytical analysis. Potentially enabling a path integral formulation (especially for $q>2$)—an interesting direction for future work. 
	
	Given the relative simplicity of $R$-PSYK$_2$, we are particularly interested in extending this work to $R$-PSYK$_{q>2}$ models. At this stage, the theoretical analysis can only rely on the path integral approach. For the ordinary SYK model, we can introduce collective fields to reduce the interaction terms to quadratic forms. For conventional fermions, exchanging two field operators merely changes the sign, but for para-particles, the commutation relations are more complicated—swapping two field operators generates additional operators, introducing further complexity. Thus, the difficulty in retaining the field operator formalism lies in the intricate commutation relations and the path integral over $\psi_{i,a}$. If we adopt a brute-force approach, expressing the Hamiltonian in terms of local fermionic creation and annihilation operators could make the system extremely complicated. If we can successfully overcome these difficulties—just as in the ordinary SYK model—we may obtain a low-energy effective theory, enabling us to study its gravitational duality. Another promising direction involves exploring matrix ensembles beyond the Gaussian unitary ensemble (GUE). These questions remain open for future research.

	\section*{Acknowledgments}
	The graphical representations of the $R$-matrix and operators $\psi_{i,a}^{\pm}$ are adapted from the work of Zhiyuan Wang and Kaden R.A. Hazzard \cite{Wang_2025}. 
	I am grateful to Yingyu Yang, Jianghui Yu, Yanyuan Li, and Professor Cheng Peng for their insightful discussions, as well as to Zhiyuan Wang, Chen-Te Ma, and Benjamin James Pethybridge for their kind comments and suggestions. T.L. acknowledges support from the National Natural Science Foundation of China (Grant No. 12175237).
	%%%%
	\appendix
	\section{Cluster Function Approach}
	\label{appendix:cluster}
	
	In this appendix, we provide a concise derivation of the cluster function approach for computing the spectral form factor (SFF).\footnote{For further details, see Appendix~A of \cite{Liao_2020}.} For a product $G=\prod_{j=1}^N (1+F(\varepsilon_j,t))$ with any function $F(\varepsilon,t)$, we begin with the expectation value
	\begin{align}
		\mathbb{E}\left(\prod_{j=1}^{N}\left[1+F\left(\varepsilon_{j},t\right)\right]\right) 
		&= \int\prod_{i=1}^{N}d\varepsilon_{i}P_{N}(\varepsilon_{1},\ldots,\varepsilon_{N})\prod_{j=1}^{N}\left[1+F\left(\varepsilon_{j},t\right)\right] \nn
		&= 1+\sum_{n=1}^{N}\frac{1}{n!}\int\mathsf{R}_{n}(\varepsilon_{1},\ldots,\varepsilon_{n})\prod_{j=1}^{n}F\left(\varepsilon_{j},t\right)d\varepsilon_{j}\ed
	\end{align}
	where we have used the fact the joint probability density $P_{N}(\varepsilon_{1},\ldots,\varepsilon_{N})$ is symmetric under the permutation of its arguments. Here, $\mathsf{R}_{n}(\varepsilon_{1},\ldots,\varepsilon_{n})$ represents the $n$-point single-particle energy level correlation function, defined as
	\begin{align}
		\mathsf{R}_{n}(\varepsilon_{1},\ldots,\varepsilon_{n}) = \frac{N!}{(N-n)!}\int d\varepsilon_{n+1}\ldots d\varepsilon_{N}P_{N}(\varepsilon_{1},\ldots,\varepsilon_{N})\ed
	\end{align}
	In the large-$N$ limit, this correlation function is given by the determinant of the kernel $\mathsf{K}(\varepsilon_i,\varepsilon_j)$
	\begin{align}
		\mathsf{R}_n(\varepsilon_{1},\varepsilon_{2},\ldots,\varepsilon_{n})=\det\left[\mathsf{K}(\varepsilon_{i},\varepsilon_{j})\right]_{i,j=1,\ldots,n}
	\end{align} 
    where
	\begin{align}
		\mathsf{K}(\varepsilon_i,\varepsilon_j) = 
		\begin{cases} 
			\frac{N}{2\pi}\sqrt{4-\varepsilon_i^2}\,\Theta(2-|\varepsilon_i|), & i=j, \\[10pt]
			\frac{N}{\pi}\frac{\sin\left[N(\varepsilon_i-\varepsilon_j)\right]}{N(\varepsilon_i-\varepsilon_j)}, & i\neq j.
		\end{cases}
	\end{align}
	Here $\Theta(x)$ is the Heaviside theta function. We then introduce the $n$-point cluster function
	\begin{align}
		\mathsf{T}_n(\varepsilon_1, \dots, \varepsilon_n) = \sum_{\mathcal{P}(n)} \mathsf{K}(\varepsilon_1, \varepsilon_2) \cdots \mathsf{K}(\varepsilon_{n-1}, \varepsilon_n) \mathsf{K}(\varepsilon_n, \varepsilon_1),
	\end{align}
	where the summation is over $(n-1)!$ cyclic permutations $\mathcal{P}(n)$ of indices $\{1,2,\ldots, n\}$. Finally we have the compact expression for large $N$
	\begin{align}\label{eq:G-t}
		\langle G \rangle = \exp\left[\sum_{n=1}^{N}\frac{(-1)^{n-1}}{n!}\bar{\mathsf{t}}_{n}\right]
	\end{align}
	with
	\begin{align}
		\bar{\mathsf{t}}_{n} = \int d\varepsilon_{1}\ldots d\varepsilon_{n}\,\mathsf{T}_{n}(\varepsilon_{1},\dots,\varepsilon_{n})\prod_{i=1}^{n}F(\varepsilon_{i},t)\ed
	\end{align}
	For both the partition function and SFF calculations considered in this work, the product of $F$ functions can be expressed as a sum
	\begin{align}\label{eq:Fexp}
		\prod_{j=1}^{n}F(\varepsilon_{j},t) = 
		\begin{cases}
			\sum_{\{\zeta\}}c_{\{\zeta\}}e^{it\sum_{j=1}^{n}\varepsilon_{j}\zeta_{j}} & \text{(SFF at infinite temperature)}, \\
			\sum_{\{\zeta\}}c'_{\{\zeta\}}e^{-\beta\sum_{j=1}^{n}\varepsilon_{j}\zeta_{j}} & \text{(Partition function)},
		\end{cases}
	\end{align}
	where $\zeta_i=-L,-L+1,\ldots,L,L\in \mathbb{N}^+$. Substituting Eq.~\eqref{eq:Fexp} into Eq.~\eqref{eq:G-t} reduces our problem to evaluating two types of integrals
	\begin{align}
		I_{n}(t\{\zeta_{i}\}) &\equiv \int d\varepsilon_{1}\ldots d\varepsilon_{n}\,\mathsf{K}(\varepsilon_{1},\varepsilon_{2})\ldots\mathsf{K}(\varepsilon_{n},\varepsilon_{1})e^{it\sum_{i=1}^{n}\varepsilon_{i}\zeta_{i}}, \\
		I_{n}^E(\beta\{\zeta_{i}\}) &\equiv \int d\varepsilon_{1}\ldots d\varepsilon_{n}\,\mathsf{K}(\varepsilon_{1},\varepsilon_{2})\ldots\mathsf{K}(\varepsilon_{n},\varepsilon_{1})e^{-\beta\sum_{i=1}^{n}\varepsilon_{i}\zeta_{i}}.
	\end{align}
	While one might expect $I_{n}^E(\{\zeta_{i}\})$ to be simply related to $I_{n}(\{\zeta_{i}\})$ by analytic continuation $t\to i\beta$, this proves nontrivial in practice. To evaluate $I_{n}(\{\zeta_{i}\})$, we employ the box approximation (detailed in Eq.~(S21-S25) of \cite{Liao_2020}), yielding
	\begin{align}\label{eq:In}
		I_{1}(\zeta_{1}t) &= N\frac{J_{1}(2|\zeta_{1}|t)}{|\zeta_{1}|t}, \nn
		I_{n \geq 2}(t\{\zeta_i\}) &= N \frac{\sin\left[\frac{\pi}{2}\left(t \sum_{j=1}^n \zeta_j\right)\right]}{\frac{\pi}{2} \left(t \sum_{j=1}^n \zeta_j\right)} \left[1 - \frac{t}{2N} s(\{\zeta_i\})\right] \Theta \left[1 - \frac{t}{2N} s(\{\zeta_i\})\right],
	\end{align}
	where
	\begin{align}
		s(\{\zeta_i\}) = \max\left\{0, \sum_{i=1}^j \zeta_i\right\}_{j=1}^{n-1} - \min\left\{0, \sum_{i=1}^j \zeta_i\right\}_{j=1}^{n-1}.
	\end{align}
	The presence of the Heaviside theta function $\Theta(x)$, defined only for real $x$, prevents a naive analytic continuation to obtain $I_{n}^E(\{\zeta_{i}\})$. Following \cite{Cotler_2017}, where the box approximation was justified for infinite temperature, we extend this approach to the high-temperature limit ($\beta \ll 1$). By neglecting the term $\left[1 - \frac{t}{2N} s(\{\zeta_i\})\right] \Theta \left[1 - \frac{t}{2N} s(\{\zeta_i\})\right]$ in Eq.~\eqref{eq:In} and substituting $t\to i\beta$, we obtain
	\begin{align}
		I_{1}^E(\beta\zeta) &= N \,_{0}F_{1}\left(2;\beta^2\zeta^{2}\right), \\
		I^E_{n\geq2}(\beta\{\zeta\}) &= N \frac{\sinh\left(\frac{\pi\beta}{2}\sum_{i=1}^{n}\zeta_{i}\right)}{\frac{\pi\beta}{2}\sum_{i=1}^{n}\zeta_{i}}.
	\end{align}
	\section{Coherent State Approach}
	\label{appdix:coherent}
	For any quantity $G = \prod_{i=1}^N \mathcal{G}(\varepsilon_i)$, we use the exact expression
	\begin{align}
		P(\varepsilon_{1},...,\varepsilon_{N}) = \frac{(N/2)^{N/2}}{N!}\sum_{i,j}\varepsilon_{\{i\}}^{\{j\}}\prod_{k=1}^{N}e^{-N\varepsilon_{k}^{2}/2}\mathsf{H}_{i_{k}-1}\left(\sqrt{\frac{N}{2}}\varepsilon_{k}\right)\mathsf{H}_{j_{k}-1}\left(\sqrt{\frac{N}{2}}\varepsilon_{k}\right),
	\end{align} 
	where $\mathsf{H}_{k}(x) = \frac{1}{\sqrt{2^k k! \sqrt{\pi}}}H_{k}(x)$ are the normalized Hermite polynomials. We then find
	\begin{align}
		\langle G \rangle &= \frac{(N/2)^{N/2}}{N!}\sum_{i,j}\varepsilon_{\{i\}}^{\{j\}}\prod_{k=1}^{N}\int d\varepsilon_{k}\, e^{-N\varepsilon_{k}^{2}/2}\mathsf{H}_{i_{k}-1}\left(\sqrt{\frac{N}{2}}\varepsilon_{k}\right)\mathsf{H}_{j_{k}-1}\left(\sqrt{\frac{N}{2}}\varepsilon_{k}\right)\prod_{i=1}^{N}\mathcal{G}(\varepsilon_i)\nn
		&= \frac{1}{N!}\sum_{i,j}\varepsilon_{\{i\}}^{\{j\}}\prod_{k=1}^{N}\int dx\, e^{-x^{2}}\mathsf{H}_{i_{k}-1}(x)\mathsf{H}_{j_{k}-1}(x) \mathcal{G}\left(x\sqrt{\frac{2}{N}}\right) \nn
		&= \det\mathsf{M}\co
	\end{align} 
	where 
	\begin{align}\label{eq:coherent-approx}
		\mathsf{M}_{ij} &= \int dx\, e^{-x^{2}}\mathsf{H}_{i-1}(x)\mathsf{H}_{j-1}(x)\mathcal{G}\left(x\sqrt{\frac{2}{N}}\right) \nn
		&= \langle i-1|\mathcal{G}\left(\hat{x}\sqrt{\frac{2}{N}}\right)|j-1\rangle.
	\end{align}
	Noting that $e^{-x^2/2}\mathsf{H}_k(x)$ is the $k$-th wavefunction of a quantum harmonic oscillator with $m = \omega = 1$, we can treat $x$ as an operator. We evaluate the large $N$ limit as
	\begin{align}\label{eq:G_trace}
		\frac{1}{N}\log\langle G \rangle = \frac{1}{N}\log\det\mathsf{M} \approx \frac{1}{N}\sum_{i=1}^{N}\log\lambda_{i},
	\end{align}
	where $\lambda_i$ are the eigenvalues of the operator $\mathcal{G}\left(\hat{x}\sqrt{\frac{2}{N}}\right)$. Note that $\mathsf{M}$ is an $N$-dimensional matrix obtained by truncating the infinite-dimensional matrix $\mathcal{G}\left(\hat{x}\sqrt{\frac{2}{N}}\right)$, so its eigenvalues are not exactly the same. However, we expect the approximation in Eq.~\eqref{eq:G_trace} to hold for large $N$.
	Using the relation $\hat{x} = \frac{1}{\sqrt{2}}(a + a^\dagger)$, where $a$ and $a^\dagger$ are the harmonic oscillator ladder operators, we employ coherent states $|\alpha\rangle$ (eigenstates of $a$) as our basis
	\begin{align}
		\langle \alpha | \alpha' \rangle &= \delta^2(\alpha - \alpha'), \quad \forall \alpha, \alpha' \in \mathbb{C}, \\
		\mathbb{I} &= \int \frac{d^2\alpha}{\pi}|\alpha\rangle\langle\alpha|.
	\end{align}
	The trace in Eq.~\eqref{eq:G_trace} can be interpreted as a truncated trace in the harmonic oscillator Hilbert space
	\begin{align}
		\text{tr}_N(\bullet) \equiv \text{tr}(\mathbb{I}_N\bullet \mathbb{I}_N) = \sum_{k=0}^{N-1}\langle k|\bullet|k\rangle \Rightarrow \frac{1}{N}\log\langle G \rangle \approx \frac{1}{N}\text{tr}_N\log\left[\mathcal{G}\left(\hat{x}\sqrt{\frac{2}{N}}\right)\right],
	\end{align}
	where $\mathbb{I}_N = \sum_{i=0}^{N-1}|i\rangle\langle i|$ is the projector onto the first $N$ states. For the coherent state basis, we restrict the trace to states with $|\alpha| \leq R$ and take the following replacements
	\begin{align}\label{eq:trN-to-trR}
		N &= \text{tr}_R\mathbb{I} = R^{2}, \\
		\text{tr}_N \hat{A} &\to \text{tr}_R(\hat{A}) = \int_{|\alpha|<R}\frac{d^2\alpha}{\pi}\langle \alpha |\hat{A}|\alpha \rangle, \\
		\hat{x}\sqrt{\frac{2}{N}} &\to \frac{1}{R}(a + a^{\dagger}).
	\end{align}
	\begin{figure}[h]
		\centering
		\includegraphics[width=0.4\textwidth]{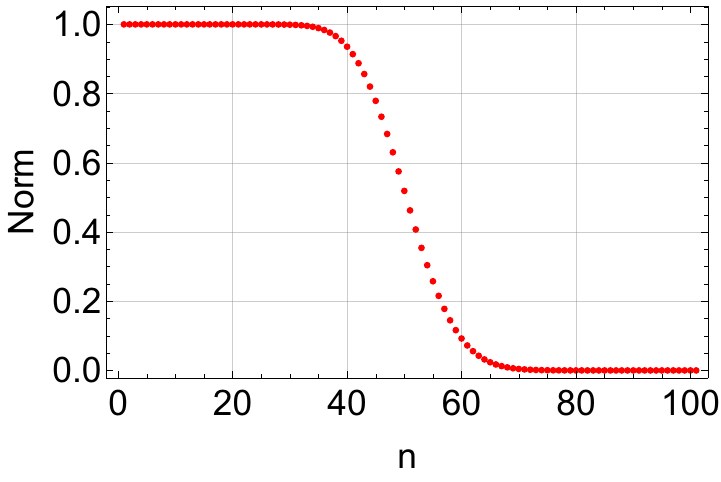}
		\includegraphics[width=0.4\textwidth]{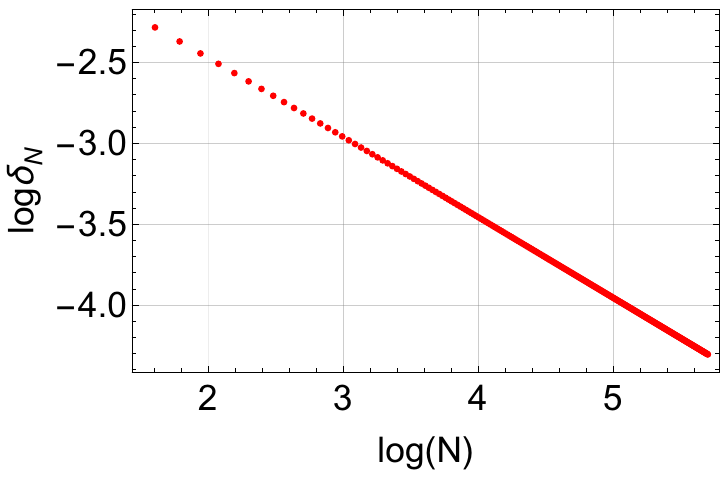}
		\caption{\textbf{Left}: Norm from Eq.~\eqref{eq:norm} with $N=50$. \textbf{Right}: Log-log plot of the error in Eq.~\eqref{eq:error}.}
		\label{fig:norm}
	\end{figure}
	\begin{figure}[h]
		\centering
		\includegraphics[width=0.4\textwidth]{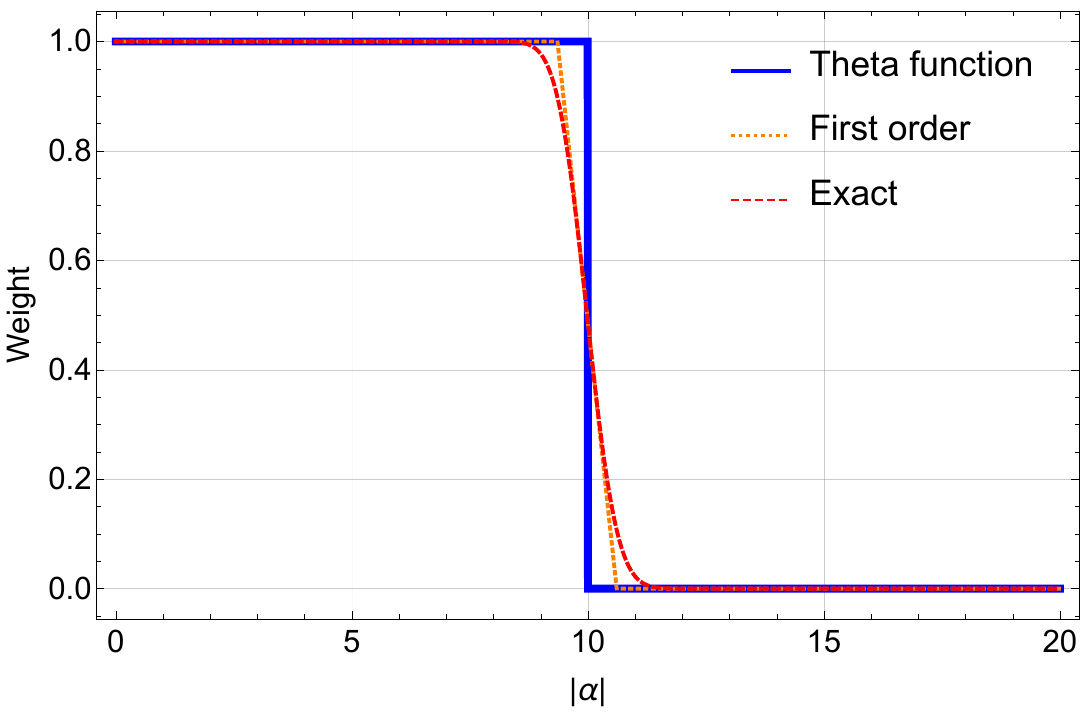}
		\includegraphics[width=0.4\textwidth]{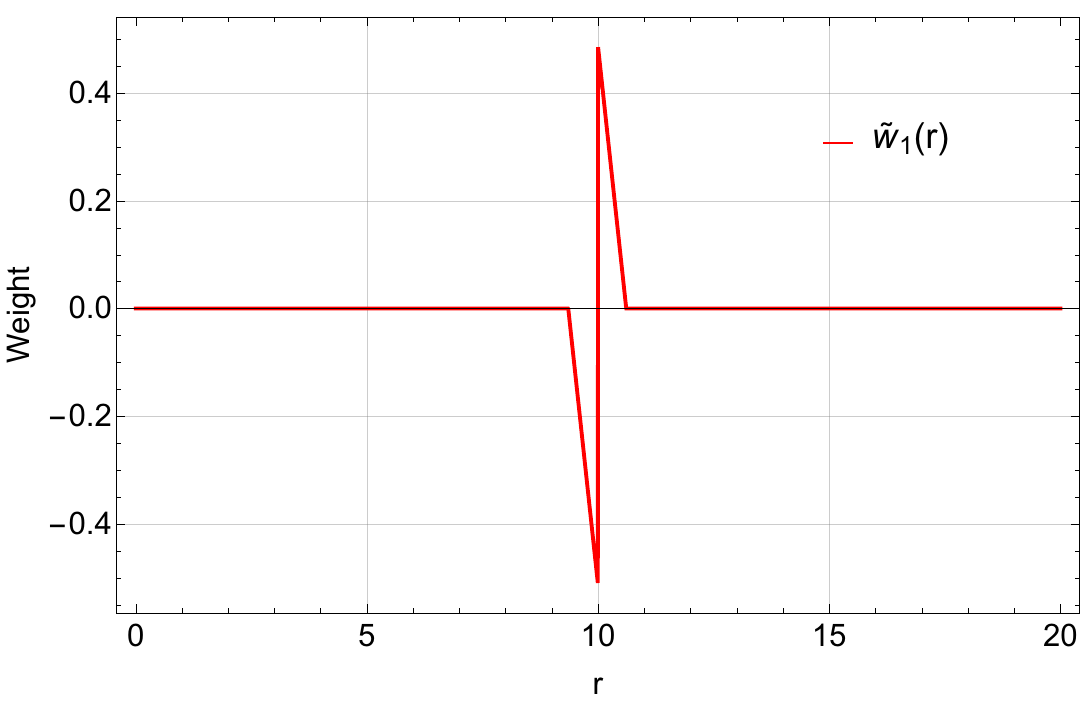}
		\caption{Leading-order and first-order approximations of the weight function.}
		\label{fig:weight}
	\end{figure}
	To quantify the difference between the two trace methods, we project $|n\rangle$ onto the coherent state basis and compute the norm $w_n$
	\begin{equation}\label{eq:norm}
		\begin{aligned}
			w_n &= \int_{|\alpha|\leq R}\frac{d^2\alpha}{\pi}|\langle n|\alpha\rangle|^2 = \int_{|\alpha|\leq R}\frac{d^2\alpha}{\pi}e^{-|\alpha|^{2}}\frac{|\alpha|^{2n}}{n!} \\
			&= R^{2}\int_{0}^{1}\int_{0}^{2\pi}\frac{u\,du\,d\theta}{\pi}e^{-u^{2}R^{2}}\frac{u^{2n}R^{2n}}{n!} = 1 - \frac{\Gamma(n+1,N)}{\Gamma(n+1)}.
		\end{aligned}
	\end{equation}
	As shown in Fig~\ref{fig:norm}, the error between the exact and approximate traces is measured by
	\begin{align}\label{eq:error}
		\delta_N &= \frac{1}{N}\sum_{n=0}^{\infty}(w_{n} - w_{n}^{\text{exact}})^{2} \sim \frac{a}{N^b}, \\
		a &\approx 0.229955, \quad b \approx 0.497012.
	\end{align}
	Here the exact norm is $w^{\text{exact}}_{n<N}=1,w^{\text{exact}}_{n\ge N}=0$. The vanishing of $\delta_N$ as $N \to \infty$ confirms the equivalence of the two traces in this limit.
	For $R \gg 1$, we approximate the trace as
	\begin{equation}\label{eq:Coherence-appro}
		\begin{aligned}
			\frac{1}{N}\log\langle G\rangle &\approx \frac{1}{R^{2}}\text{tr}_{R}\log\left[\mathcal{G}\left(\frac{a + a^{\dagger}}{R}\right)\right] \\
			&\approx \int_{0}^{1}\int_{0}^{2\pi}\frac{u\,du\,d\theta}{\pi}\log\left[\mathcal{G}(2u\cos\theta)\right],
		\end{aligned}
	\end{equation}
	where we used polar coordinates $\alpha = Rue^{i\theta}$. For operators of the form $A = e^{\lambda(a + a^\dagger)}$, which satisfy $A|\alpha\rangle = e^{\lambda(\alpha^* + \alpha - 1/2)}|\alpha\rangle$, we have
	\begin{align}
		\text{tr}_{N}(A) &= \sum_{n=0}^{N-1}\int\frac{d^2\alpha}{\pi}e^{-|\alpha|^{2}}\frac{|\alpha|^{2n}}{n!}A(\alpha,\alpha^{*}).
	\end{align}
	The weight function approximation can be improved by expanding near $r = R = \sqrt{N}$
	\begin{align}
		w(r) &= \begin{cases}
			1, & r < r_{a}, \\
			w_{p}(r), & r_{a} \leq r \leq r_{b}, \\
			0, & r > r_{b},
		\end{cases}
	\end{align}
	where $w_{p}(r) = \sum_{j=0}^{p}c_{j}(r - R)^{j}$. The first-order approximation gives
	\begin{equation}\label{eq:first-order}
		\begin{aligned}
			w_{1}(r) &= \frac{\Gamma(N,N)}{\Gamma(N)} - \frac{2e^{-N}N^{N+\frac{1}{2}}(r - \sqrt{N})}{\Gamma(N+1)}, \\
			r_{a} &= \sqrt{N} - \frac{N^{-N-\frac{1}{2}}e^{N}(\Gamma(N)^2 - \Gamma(N)\Gamma(N,N))}{2\Gamma(N)}, \\
			r_b &= \sqrt{N} + \frac{N^{-N-\frac{1}{2}}e^{N}\Gamma(N+1)\Gamma(N,N)}{2\Gamma(N)}.
		\end{aligned}
	\end{equation}
	In the large $N$ limit, we find
	\begin{align}
		\lim_{N\to \infty} \frac{\Gamma(N,N)}{\Gamma(N)} &= \frac{1}{2}, \\
		\lim_{N\to \infty}\frac{\sqrt{N} - r_a}{r_b - \sqrt{N}} &= 1,
	\end{align}
	which suggests the first-order approximation
	\begin{align}
		w(r) \approx \Theta(\sqrt{N} - r) + \widetilde{w}_1(r).
	\end{align}
	The leading-order and first-order approximations are compared in Fig.~\ref{fig:weight}.
	\section{Calculation of $B_0$ and $C_0$ via Transfer Matrix}
	\label{appdix:transferT}
	The coefficient $C_0$ is defined as
	\begin{align}
		C_{0} = \sum_{n=2}^{N}\frac{(-1)^{n}}{n2^{n+1}}\sum_{\substack{\sum\zeta_{i}=0}}\mathsf{g}(\{\zeta_{i}\})s(\{\zeta_{i}\}).
	\end{align}
	Consider paths generated by displacements $\{\zeta_i\}$ where $x_j = \sum_{i=1}^j \zeta_i$ with $x_0 = 0$. Each path has displacements $\zeta_{j+1} \in \{0, \pm1, \pm2, \ldots, \pm L\}$ weighted by $\mathsf{g}_{|\zeta_{j+1}|}$. The span factor $s(\{\zeta_i\}) = \max(\{x\}) - \min(\{x\})$ accounts for the vertical extent of each path.
	Using the symmetry $\mathsf{g}_k = \mathsf{g}_{-k}$, we rewrite $C_0$ as
	\begin{align}
		C_0 = \sum_{n=2}^{N}\frac{(-1)^{n}}{n2^{n}} \sum_{k=1}^{\lfloor Ln/2 \rfloor} kN_{n,k}^{(0)},
	\end{align}
	where $N_{n,k}^{(0)}$ enumerates weighted paths of length $n$ with maximum displacement $k$
	\begin{align}
		N_{n,k}^{(0)} = \sum_{\substack{\text{Length}=n \\ \max|\{x\}|=k}} \text{weight}\{x\}.
	\end{align}
	The transfer matrix $T^{\le k}$ of dimension $(k+\lfloor nL/2 \rfloor +1)$ has elements
	\begin{align}
		T^{\le k}_{ij} = \mathsf{g}_{|i-j|}, \quad i,j = -\lfloor nL/2 \rfloor, \ldots, k,
	\end{align}
	forming a symmetric Toeplitz matrix with bandwidth $L$
	\begin{align}
		T^{\le k} = 
		\begin{bmatrix} 
			\mathsf{g}_0 & \mathsf{g}_1 & \cdots & \mathsf{g}_L & 0 & \cdots \\
			\mathsf{g}_1 & \mathsf{g}_0 & \mathsf{g}_1 & \ddots & \ddots & \\
			\vdots & \ddots & \ddots & \ddots & \ddots & \mathsf{g}_L \\
			\mathsf{g}_L & \ddots & \ddots & \ddots & \ddots & 0 \\
			0 & \ddots & \ddots & \ddots & \ddots & \vdots \\
			\vdots & & \mathsf{g}_L & \cdots & \mathsf{g}_1 & \mathsf{g}_0
		\end{bmatrix}.
	\end{align}
	Path counting relations are obtained through the transfer matrix
	\begin{align}
		\sum_{j=0}^{k} N_{n,j}^{(0)} = \bra{0}\left(T^{\le k}\right)^n\ket{0},
	\end{align}
	yielding the fundamental identity
	\begin{align}
		\sum_{k=0}^{\lfloor nL/2\rfloor} kN_{n,k}^{(0)} = (\lfloor nL/2\rfloor+1)\bra{0}T^n\ket{0} - \sum_{k=0}^{\lfloor nL/2\rfloor} \bra{0}\left(T^{\le k}\right)^n\ket{0},
	\end{align}
	where $T \equiv T^{\le \lfloor nL/2\rfloor}$.
	The final expressions for the coefficients are
	\begin{align}
		C_{0} &= \sum_{n=2}^{N}\frac{(-1)^{n}}{n2^{n}}\left[(\lfloor nL/2\rfloor+1)\bra{0}T^n\ket{0} - \sum_{k=0}^{\lfloor nL/2\rfloor}\bra{0}\left(T^{\le k}\right)^n\ket{0}\right], \\
		B_{0} &= \sum_{n=2}^{N}\frac{(-1)^{n-1}}{n2^n}\bra{0}T^n\ket{0}.
	\end{align}
	For numerical implementation, we shift indices to $1,\ldots,\lfloor nL/2 \rfloor+1+k$ and represent $\ket{x}$ as unit vectors with 1 at position $(x+\lfloor nL/2 \rfloor+1)$.

	\bibliographystyle{JHEP}
	\bibliography{ref}
\end{document}